\documentclass[aps,prd,nofootinbib,two column]{revtex4-2}
\usepackage{graphicx}
\usepackage{subcaption}
\usepackage{natbib}
\usepackage{mathtools}
\usepackage{slashed}
\usepackage{url}
\usepackage[normalem]{ulem}
\usepackage{enumitem}
\usepackage{mathrsfs}
\usepackage{float}
\usepackage{multirow}

\usepackage[colorlinks=true, linkcolor=blue, citecolor=red, urlcolor=magenta]{hyperref}
\newcommand{\orcid}[1]{\href{https://orcid.org/#1}{\includegraphics[width=8pt]
		{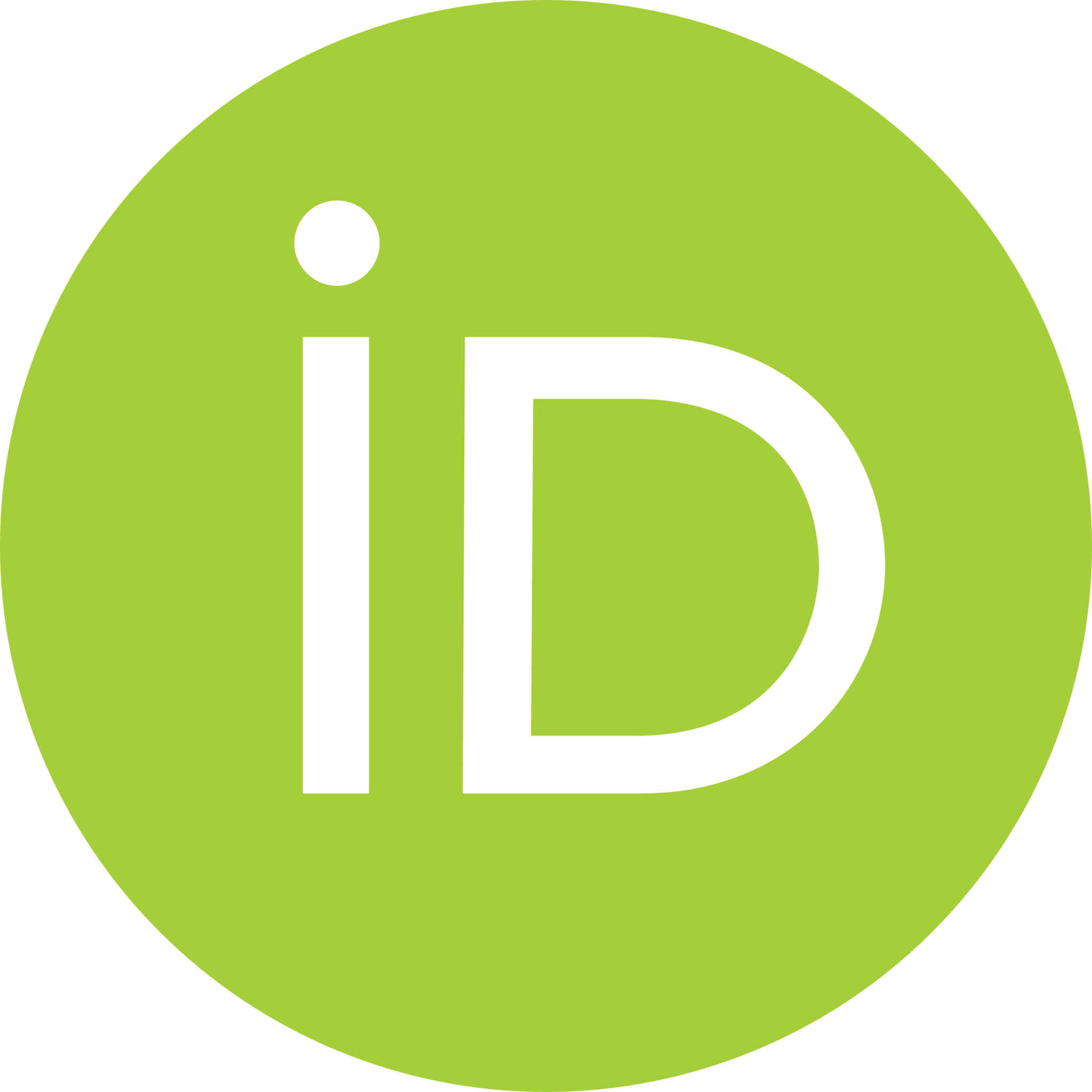}}}

\begin{document}
\title{Spectra and elliptic flow of light hadrons in an expanding fire-cylinder model for the RHIC Beam Energy Scan}
\author{Anand Rai\orcid{https://orcid.org/0009-0005-4761-7918}}
\email{ranand@iitbhilai.ac.in}
\affiliation{Department of Physics, Indian Institute of Technology Bhilai, Kutelabhata, Durg, 491002, Chhattisgarh, India}   
	
\author{Ashutosh Dwibedi\orcid{https://orcid.org/0009-0004-1568-2806}}
\email{ashutoshdwibedi92@gmail.com}	
\affiliation{Department of Physics, Indian Institute of Technology Bhilai, Kutelabhata, Durg, 491002, Chhattisgarh, India}

\author{Sabyasachi Ghosh\orcid{https://orcid.org/0000-0003-1212-824X}}
\email{sabya@iitbhilai.ac.in}
\affiliation{Department of Physics, Indian Institute of Technology Bhilai, Kutelabhata, Durg, 491002, Chhattisgarh, India}

\begin{abstract}
We investigate the transverse momentum spectra ($p_T$) and elliptic flow ($v_2$) of $\pi^{\pm}$, $K^{\pm}$, $p$, and $\bar{p}$ produced in mid-central Au+Au collisions at $\sqrt{s_{\rm NN}} = 7.7$, 11.5, 19.6, 27, and 39 GeV in the Beam Energy Scan (BES) Program at the Relativistic Heavy Ion Collider (RHIC). The analysis is carried out within an expanding elliptic fire-cylinder model that incorporates longitudinal expansion and anisotropic transverse flow. Particle production at kinetic freeze-out is obtained using a local equilibrium distribution function with a blast-wave–like fluid velocity profile derived from the expansion dynamics of the elliptic fire-cylinder. The model parameters governing the collective expansion are first constrained by fitting the mid-rapidity $p_T$ spectra of $\pi^{\pm}$ and are then applied, without further adjustment, to $K^{\pm}$, $p$, and $\bar{p}$. The model provides a consistent description of the $p_T$ spectra and reproduces the qualitative behavior of the elliptic flow for all considered particle species.
\end{abstract}

\maketitle


\section{Introduction}
\label{sec:intro}
Relativistic heavy-ion collisions (HICs) provide a unique experimental environment to study strongly interacting matter at extreme temperatures and energy densities. Under such conditions, Quantum Chromodynamics (QCD) predicts a transition from confined hadronic matter to a deconfined quark–gluon plasma (QGP)~\cite{Shuryak:1980tp,Rischke:2003mt,Shuryak:2004cy,Busza:2018rrf,Harris:1996zx,Harris:2023tti,braun2007quest,Andronic:2017pug,Florkowski:2010zz,Letessier:2022fax}. Extensive experimental programs at the Relativistic Heavy Ion Collider (RHIC)~\cite{STAR:2017sal,Chen:2024aom} and the Large Hadron Collider (LHC)~\cite{Roland:2014jsa,Foka:2016vta} have provided compelling evidence that the medium created in these collisions behaves as a strongly coupled, nearly perfect fluid. In particular, the observed small shear viscosity to entropy density ratio and the presence of pronounced collective flow are consistent with hydrodynamic evolution of a QGP formed in the hot and dense initial stage of the collision~\cite{Shuryak:2003xe,Romatschke:2007mq}. Among the most informative observables are the transverse momentum ($p_T$) spectra of identified hadrons and the anisotropic flow coefficients $v_n$, especially the elliptic flow~\cite{Ollitrault:1992bk,Voloshin:2008dg}. The $p_T$ spectra carry information about the collective radial expansion and the kinetic freeze-out conditions of the system. The elliptic flow reflects the conversion of the initial spatial anisotropy of the collision zone into momentum-space anisotropy driven by pressure gradients~\cite{Kolb:2003dz,Huovinen:2001cy,Retinskaya:2013gca}. The observed magnitude and mass ordering of $v_2(p_T)$ provide strong evidence for early thermalization and the development of collective behavior in the medium~\cite{STAR:2004jwm,STAR:2007afq}.

Relativistic hydrodynamics has been highly successful in describing collective phenomena such as particle spectra and flow over a wide range of collision energies and system sizes~\cite{Kolb:2003dz,Heinz:2013th,Gale:2013da,Romatschke:2017ejr,De:2022yxq,Ali:2024zvp,10.1093/ptep/pts014}. Owing to the numerical complexity of full hydrodynamic simulations, simplified models inspired by hydrodynamics—commonly referred to as blast-wave (BW) models—are widely used to describe particle spectra and collective flow~\cite{PhysRevLett.42.880,Schnedermann:1993ws,Broniowski:2001we,Broniowski:2001uk,Huovinen:2001cy,Florkowski:2004tn,STAR:2001ksn,Retiere:2003kf,Tang:2008ud,He:2010vw,Sun:2014rda,Melo:2015wpa,Tomasik:2024uuq,Cimerman:2017lmm,Melo:2019mpn,FLORKOWSKI2025100249,Harabasz:2020sei,Harabasz:2022rdt,Drogosz:2025vdq}. In these models, the medium expansion is characterized by a parametrized collective velocity field and a freeze-out hypersurface~\cite{Heinz:2004qz}. The commonly used BW parametrization assumes radially increasing transverse flow together with boost-invariant longitudinal expansion and freeze-out at fixed proper time, which provides a good description of transverse momentum spectra at mid-rapidity~\cite{Schnedermann:1993ws,Bjorken:1982qr}. However, this description becomes inadequate for low-energy and peripheral HICs, where collective flow develops predominantly along the reaction plane. To address this, azimuthal anisotropy in the transverse flow and spatial emission geometry has been incorporated into BW models, enabling successful descriptions of elliptic flow~\cite{Huovinen:2001cy,STAR:2001ksn,Retiere:2003kf,He:2010vw,Cimerman:2017lmm,Sun:2014rda,Tomasik:2024uuq}. In addition, spherical or spheroidal geometries with instantaneous freeze-out in laboratory time~\cite{Harabasz:2020sei,Harabasz:2022rdt} and viscous extensions including non-equilibrium corrections to the distribution function~\cite{Teaney:2003kp,Jaiswal:2015saa,Yang:2016rnw,Yang:2018ghi,Yang:2020oig,Yang:2022ixy,Yang:2023apw} have been proposed to improve the description of experimental observables.

In this work, we employ the blast-wave–like expanding fire-cylinder model introduced in Ref.~\cite{Gossiaux:2011ea} to describe the transverse momentum spectra and elliptic flow of protons, pions, and kaons in Au+Au collisions at RHIC energies, $\sqrt{s_{\rm NN}}=7.7$–$39$ GeV, for the mid-central collisions. In off-central HICs, the initial transverse geometry of the medium is anisotropic and can be approximated by an ellipse in the transverse plane. The subsequent evolution of the system involves anisotropic transverse expansion together with longitudinal expansion. The transverse size of the medium is governed by the time evolution of the major axis $a$ (along the $y$ direction) and the minor axis $b$ (along the $x$ direction), while the longitudinal extent is characterized by $z_{B}$.
To parametrize the transverse expansion, or equivalently the time dependence of $a$ and $b$, we follow the geometric parametrization of the bulk medium developed in Refs.~\cite{Rapp:1999us,Rapp:1999zw,Rapp:2000pe,Turbide:2003si,vanHees:2005wb,vanHees:2006ng,vanHees:2007th,vanHees:2011vb,Gossiaux:2011ea,vanHees:2014ida,Rapp:2014hha}, originally applied to thermal electromagnetic emission and heavy-quark diffusion. This approach of parameterizing the bulk medium expansion has never been used systematically for fitting transverse momentum spectra and elliptic flow of $\pi$, $K$, and $p$ across different beam energies, which is attempted in the present article. Our fits involve eight free parameters: the chemical potential $\mu$ and temperature $T$ at kinetic freeze-out, five parameters ($A$, $B$, $v_{\infty}$, $\Delta v$, and $v_{0}$) that characterize the time evolution of the longitudinal and transverse dimensions of the elliptic fire-cylinder, and the freeze-out time $t_{f}$. The parameters $A$ and $v_{\infty}$ describe a uniform expansion of the elliptic cross section. For a realistic off-central collision, however, the pressure gradient along the reaction plane is larger than that in the perpendicular direction, leading to a faster expansion along the direction of the impact parameter; this anisotropy in the transverse expansion rate is quantified by the parameters $B$ and $\Delta v$. The longitudinal expansion is modeled using a longitudinal expansion velocity $v_{0}$, incorporating an initial longitudinal size $z_{0}$, which can be sizable at low collision energies. The resulting fluid velocity profile is constructed from the transverse and longitudinal expansion rates of the evolving elliptic fire-cylinder. Using the obtained fluid velocity profile and the Cooper–Frye freeze-out prescription~\cite{Cooper:1974mv}, we calculate the rapidity distributions, transverse momentum spectra, and elliptic flow of light hadrons, namely $\pi^{\pm}$, $K^{\pm}$, and $p(\bar{p})$. The transverse momentum spectra and elliptic flow have been compared with the experimental results to obtain suitable values of the parameters describing the evolution of the elliptic fire-cylinder.

The paper is organized as follows. In Sec.~\ref{sec:formalism}, we present the formalism of the expanding elliptic fire-cylinder model and the expressions for particle spectra and elliptic flow. Section~\ref{sec:results} describes the fitting procedure and discusses the resulting transverse momentum spectra and elliptic flow. Finally, Sec.~\ref{sec:sum} summarizes the main findings and outlines possible directions for future work.


\section{Formalism}
\label{sec:formalism}
The theoretical calculation of the important observables in HICs--the light hadron spectra and flow requires the description of the expanding medium formed in HIC experiments. In this section, we have described the expanding elliptic fire-cylinder model used to characterize the space-time evolution of the hot and dense matter formed in relativistic HICs. The key ingredients of the model, including parameterization of the fluid velocity and subsequent evaluation of the particle spectra and flow, are discussed. 

\subsection{The expanding elliptic fire-cylinder}
\label{subsec:fireball}

We will describe the space-time evolution of the expanding medium, which will be essentially an expanding fire-cylinder with an elliptic cross section. The total volume of the cylinder at time $t$ is parametrized as~\cite{Gossiaux:2011ea},
\begin{equation}
	V(t) = 2\pi\, a(t)\, b(t)\, z_{B}(t),
	\label{eq:volume}
\end{equation}
where $a(t)$, $b(t)$, and $z_{B}(t)$ denote the time-dependent boundary for the transverse semi-axes along
the $y$, $x$ and the longitudinal axis along the $z$ directions, respectively. 
The transverse and longitudinal scale parameters evolve according to~\cite{Gossiaux:2011ea}
\begin{align}
	a(t) &= a_0
	+ v_\infty\!\left[t - \frac{1 - e^{-A t}}{A}\right]
	- \Delta v\!\left[t - \frac{1 - e^{-B t}}{B}\right],
	\label{eq:a_t} \\
	b(t) &= b_0
	+ v_\infty\!\left[t - \frac{1 - e^{-A t}}{A}\right]
	+ \Delta v\!\left[t - \frac{1 - e^{-B t}}{B}\right],
	\label{eq:b_t}\\
	z_{B}(t)&= z_0 + v_0 t,
\end{align}
where $a_0$, $b_0$, and $z_{0}$ fix the initial transverse and longitudinal geometry of the overlap region. The initial size in the longitudinal as well as transverse directions is determined from the collision specifications, i.e., the beam energy and centrality. For each colliding beam, the energy per nucleon is given by
$E_{\text{beam}} = \frac{\sqrt{s_{\rm NN}}}{2}$ which determines the Lorentz factor
$\gamma = \frac{E_{\text{beam}}}{m_{\rm N}}$,
where $m_{\rm N}$ is the nucleon mass. The initial longitudinal size of the fire-cylinder can now be obtained as $z_0 = \frac{a_{0}}{\gamma}$, where $a_{0}=\sqrt{R^{2}-b_{0}^{2}}$ with $R$ being the radius of the colliding nuclei and $b_{0}$ being half of the impact parameter. Our choice of describing the Au$+$Au collisions with centrality class $40--50$\%  translates to initial sizes $b_{0}=3.5$ fm and $a_{0}=6.06$ fm. 
The quantity $v_\infty$ denotes the asymptotic transverse expansion velocity, while $\Delta v$ quantifies the strength of the anisotropic flow component. The parameters $A$ and $B$ control the time scales for the buildup of isotropic and anisotropic collective flow, respectively. $v_0$ is the effective longitudinal expansion velocity. 
The transverse expansion rates are given by
\begin{align}
	\dot{a}=\frac{d a}{d t} &= v_\infty\!\left(1 - e^{-A t}\right)
	- \Delta v\!\left(1 - e^{-B t}\right), \label{eq:adot} \\
	\dot{b}=\frac{d b}{d t} &= v_\infty\!\left(1 - e^{-A t}\right)
	+ \Delta v\!\left(1 - e^{-B t}\right). \label{eq:bdot}
\end{align}
We will express the fluid four-velocity using the expansion rates given above. Before writing the fluid profile, we briefly describe the coordinate system we will use in this work. In the transverse plane, we use the confocal elliptic coordinate system in which we have $x=\lambda \sinh\xi \cos\theta$ and $y=\lambda \cosh\xi \sin\theta$, where $\xi\geq 0$ and $0\leq \theta \leq 2\pi$. For a given $\lambda$ (corresponding to the focal length), these define a confocal ellipse (for constant $\xi$) and a hyperbola (for constant $\theta$) having their major axis aligned with the $y$ axis. The system size in the transverse plane at time $t$ is given by,
\begin{equation}
	\frac{x^2}{b^2(t)} +
	\frac{y^2}{a^2(t)} \leq 1 .
	\label{eq:ellipse}
\end{equation}
At a fixed time $t$ we set-up an elliptic coordinate system with $\lambda(t)=\sqrt{a(t)^{2}-b(t)^{2}}$ and $0\leq \xi \leq \xi_{\rm max}(t)$, where $\xi_{\rm max}(t)=\sinh^{-1}\frac{b(t)}{\lambda(t)}$. The jacobian for the coordinate transformation $(x,y)\rightarrow(\xi, \theta)$ is given by, $J(\xi,\theta) =\lambda^{2}(t)\left[\cosh^2 \xi\cos^2\theta+ \sinh^2 \xi \sin^2\theta\right]$.

The fluid four-velocity field is written as~\cite{Heinz:2004qz},
\begin{equation}
	u^\mu =
	\gamma_T \cosh\eta_L
	\left(
	1,
	\frac{v_x}{\cosh\eta_L},
	\frac{v_y}{\cosh\eta_L},
	\tanh\eta_L
	\right),
	\label{eq:four_velocity}
\end{equation}
where $\eta_L$ denotes the longitudinal fluid rapidity and $\gamma_{T}=\frac{1}{\sqrt{1-v_{x}^{2}-v_{y}^{2}}}$ is the Lorentz gamma factor associated with transverse velocity. The transverse and longitudinal velocity components at the boundary (surface) of the elliptic fire-cylinder are obtained as
\begin{align}
	v^{B}_x = \dot{b} \cos\theta, v^{B}_y =\dot{a}\sin\theta, v^{B}_{z}=v_{0}~.   \label{eq:Bv}
\end{align}
Assuming the velocity increases as we move towards the boundary, we take the following ansatz for the fluid profile~\cite{vanHees:2005wb,Gossiaux:2011ea}
\begin{align}
	v_x = \frac{r}{r_B}\, \dot{b}\, \cos\theta,~	v_y = \frac{r}{r_B}\, \dot{a}\, \sin\theta,~ v_{z}=\frac{z }{z_{B}}v_{0}.  \label{eq:v}
\end{align}
The radial lengths are defined as, $r=\sqrt{x^2+y^2}\leq r_{B}$ and $r_{B}=\sqrt{x_{B}^2+y_{B}^2}$ where the coordinates at the transverse boundary is $x_{B}=\lambda(t)\sinh\xi_{max}(t)\cos\theta$ and $y_{B}=\lambda(t)\cosh\xi_{max}(t)\sin\theta$. The longitudinal velocity is written in the same spirit as that of transverse ones with the longitudinal boundary $z_{B}(t)=z_{0}+v_{0}t$. We notice that in the limit $z_{0}=0$ we get back the Bjorken flow~\cite{Bjorken:1982qr} $v_{z}=\tanh \eta_{L} \rightarrow\frac{z}{t}$ for the region $-z_{B}\leq z\leq z_{B}$.

The invariant momentum distribution is obtained using the Cooper--Frye prescription~\cite{Cooper:1974mv},
\begin{equation}
	E\frac{d^3N}{d^3\vec{p}}=\frac{g}{(2\pi)^3} \int_{\sigma} f(x,p)~ p^\mu  d\sigma_\mu,
	\label{eq:cooper_frye1}
\end{equation}
where $f$ is the single particle distribution function, $g$ is the associated spin degeneracy factor and $\sigma$ denotes the kinetic freeze-out hyper surface. We choose a constant lab time freeze-out at time $t=t_{f}$ which corresponds to $d\sigma_\mu = (d^3x, \mathbf{0})$. Eq.~\eqref{eq:cooper_frye1} now reduces to
\begin{equation}
	E\frac{d^3N}{d^3\vec{p}}=\frac{g}{(2\pi)^3}\int E f(x,p)\, dx\,dy\,dz .\label{eq:cooper_frye2}
\end{equation}
For our purpose we use the local thermal distribution to be the Bose--Einstein or Fermi--Dirac form depend on the particle type $f(x,p) =\left[\exp\!\left(\dfrac{u_{\mu}p^{\mu} - \mu}{T}\right) \mp 1\right]^{-1}$, where $p^\mu =\left(m_T \cosh y_{p}, p_T \cos\phi_p, p_T \sin\phi_p, m_T \sinh y_{p} \right)$ with $p_{T}=\sqrt{p_{x}^{2}+p_{y}^{2}}$, $m_{T}^{2}=p_{T}^{2}+m^{2}$, $y_{p}=\frac{1}{2}\ln \frac{E+p_{z}}{E-p_{z}}$ and  $\phi_{p}$ is the momentum space azimuthal angle. Transforming to elliptic coordinates $(\xi,\theta)$ and using the identity $\frac{d^{3}\vec{p}}{E}=p_{T}dp_{T}dy_{p}d\phi_{p}$, Eq.~\eqref{eq:cooper_frye2} becomes
\begin{equation}
	\begin{aligned}
		&\frac{1}{2\pi p_T}\frac{d^3N}{dp_T dy_{p}d\phi_{p}}=\frac{g}{(2\pi)^4}\int f\left(\frac{u_{\mu}p^{\mu}-\mu}{T_{\rm kin}}\right)\, \\
		& \qquad \qquad \qquad \times  m_T \cosh y_{p}\,
		|J(\xi,\theta)|\,
		d\xi\, d\theta\,  dz ,
	\end{aligned}
	\label{eq:spectrum1}
\end{equation}
where we have $u^{\mu}p_{\mu}=\gamma_{T}(m_{T}\cosh(\eta_{L}-y_{p})-\vec{v}_{T}\cdot \vec{p}_{T})$ and $T(t=t_{f})=T_{\rm kin}$. If one approximates the distribution with the Boltzmann distribution, the above integral simplifies to
\begin{widetext}
	\begin{eqnarray}
		\frac{1}{2\pi p_T}\frac{d^3N}{dp_T dy_{p} d{\phi_{p}}}&=&\frac{g~e^{\mu/T_{\rm kin}}}{(2\pi)^4}  \int_{0}^{\xi_{f}} d\xi  \int_{0}^{2\pi} d\theta~ \lambda^{2}_{f}~ (\cosh^{2}\xi\cos^{2}\theta+\sinh^{2}\xi \sin^{2}\theta) ~e^{\left(\gamma_{T}p_{T}\frac{r}{r_{B}}(\dot{b}\cos\theta\cos\phi_{p}+\dot{a}\sin\theta\sin\phi_{p})\right)/T_{\rm kin}}  \nonumber\\
		&\times& \int_{-z_{f}}^{z_{f}} m_T \cosh y_{p}~ e^{-\gamma_{T}m_{T}\cosh(\eta_{L}-y_{p})/T_{\rm kin}}
		dz ,	\label{eq:spectrum2}
	\end{eqnarray}
	
	where $\lambda_{f}=\lambda(t_{f})$ is the focal length of the boundary ellipse at freeze-out, $\xi_{f}=\xi_{\rm max}(t_{f})$ parameterize the locus of the elliptic boundary at freeze-out, the velocity at freeze-out hyper surface are given by $v_{x}=\frac{r}{r_{B}}\dot{b}(t_{f})\cos \theta$ and $v_{y}=\frac{r}{r_{B}}\dot{a}(t_{f})\sin \theta$ with $\frac{r}{r_{B}}=\sqrt{\frac{\sinh^{2}\xi \cos^{2}\theta+\cosh^{2}\xi \sin^{2}\theta}{\sinh^{2}\xi_{f} \cos^{2}\theta+\cosh^{2}\xi_{f} \sin^{2}\theta}}$.
\end{widetext}
 
The elliptic flow coefficient is defined as \cite{Kolb:2003dz}
\begin{equation}
	v_2(y_{p},p_T) =
	\frac{
		\displaystyle
		\int d\phi_p \cos(2\phi_p)\,
		\frac{d^3N}{dp_T\,dy_{p}\,d\phi_p}
	}{
		\displaystyle
		\int d\phi_p
		\frac{d^3N}{dp_T\,dy_{p}\,d\phi_p}
	}.
	\label{eq:v2}
\end{equation}


\section{Results and Discussion}
\label{sec:results}
\subsection{Momentum spectra}
In this section, we describe the results of spectra and flow based on Eqs.~\eqref{eq:spectrum1} and \eqref{eq:v2} formulated in Sec.~\eqref{sec:formalism}.
All our model calculations correspond to mid-central collisions. The experimental data employed for comparison encompass a range of centrality classes, depending on their availability; such distinctions are explicitly indicated wherever relevant.

\begin{figure}[H]
	\centering
	\begin{subfigure}{0.45\textwidth}
		\includegraphics[width=\linewidth]{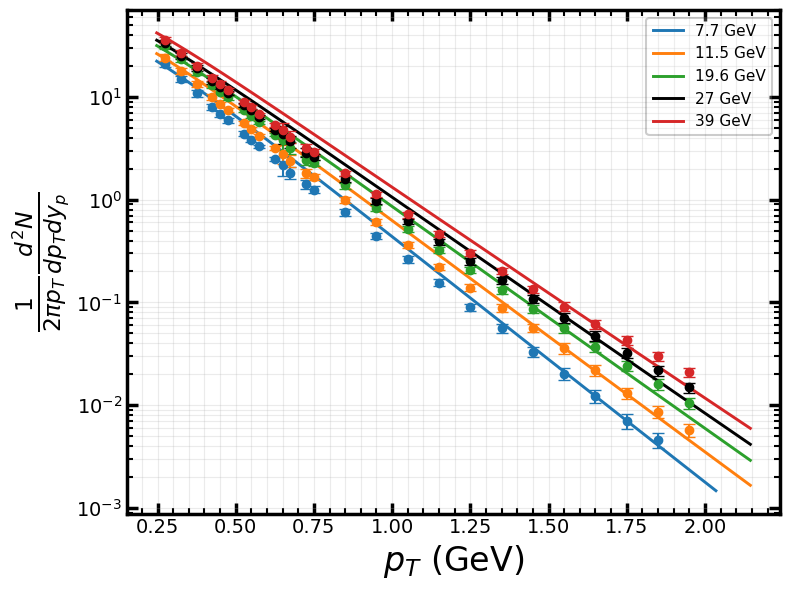}
		\caption { $p_T$ spectra of $\pi^{+}$ at mid-rapidity compared with STAR collaboration data~\cite{STAR:2017sal} for the centrality class ($40$--$50\%$).}
	\end{subfigure}
	\hfill
	\begin{subfigure}{0.45\textwidth}
		\includegraphics[width=\linewidth]{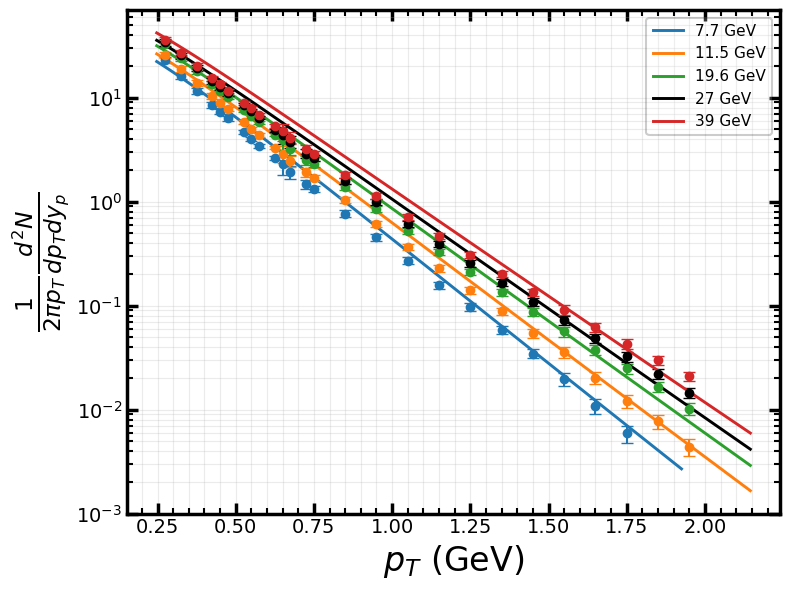}
		\caption{$p_T$ spectra of $\pi^{-}$ at mid-rapidity compared with STAR collaboration data~\cite{STAR:2017sal} for the centrality class ($40$--$50\%$).}
	\end{subfigure}
	\caption{(Color online) Calculated $p_T$ spectra of pions obtained using the fitted expansion parameters of Table~\ref{tab:model_params}.}
	\label{fig:pion-results}
\end{figure}

\begin{figure}[H]
	\centering
	\includegraphics[width=1\linewidth]{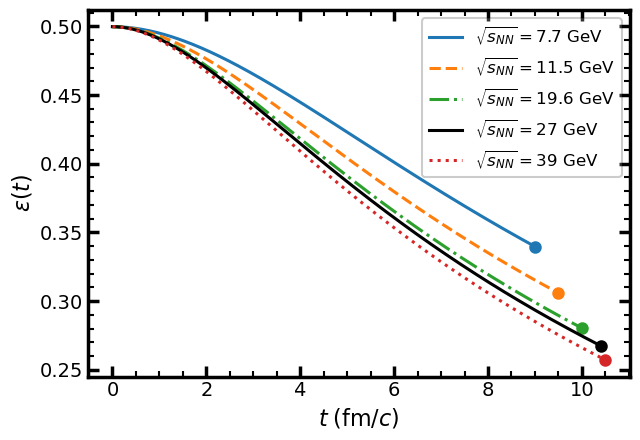}
	\caption{(Color online) Time evolution of spatial eccentricity $\epsilon(t)$ of the expanding elliptic fire-cylinder for different collision energies $\sqrt{s_{\rm NN}}$. End points denote the freeze-out times for different $\sqrt{s_{\rm NN}}$.}
	\label{fig:spacial eccentricity}
\end{figure}

The model parameters governing the collective expansion, namely the asymptotic transverse velocity $v_\infty$, the anisotropic velocity component $\Delta v$, the longitudinal velocity $v_{0}$, and the flow buildup parameters $A$ and $B$, are determined by fitting the transverse-momentum spectra of pions at mid-rapidity
($y_p = 0$) for several values of the beam energy $\sqrt{s_{\rm NN}}=7.7, 11.5, 19.6, 27, 39$ GeV. We take the pion chemical potential to be zero, i.e., $\mu_{\pi}=0$. The initial transverse dimensions $a_0$ and $b_0$ are kept fixed for all collision energies, reflecting the approximately energy-independent nuclear overlap geometry for a given centrality class, whereas $z_{0}$ is taken as a function of $\sqrt{s_{\rm NN}}$ as discussed in Sec.~\ref{subsec:fireball}. Fig.~\ref{fig:pion-results} illustrates this fit with the STAR collaboration data set~\cite{STAR:2017sal}. Same set of collective expansion parameters and a single kinetic freeze-out temperature provide a good fit for both $\pi^{+}$ and $\pi^{-}$, which justifies the assumption $\mu_{\pi^{+}}=\mu_{\pi^{-}}=0$.
Table~\ref{tab:model_params} contains the model parameters calibrated with the help of experimental $p_{T}$ spectra of the pion at mid-rapidity. In Table~\ref{tab:vel}, we compare the average surface velocity obtained in the present model with that of the STAR collaboration blast wave analysis~\cite{STAR:2017sal}. To obtain the average velocity in the transverse plane boundary at kinetic freeze-out we use Eq.~\eqref{eq:Bv} and take the angular average as follows
	\begin{eqnarray}
		&& \langle v_{B}(t_{f}) \rangle =\sqrt{\frac{1}{2\pi}\int_{0}^{2\pi} \left((v_{x}^{B})^{2} + (v_{y}^{B})^{2}\right)~ d \theta}\label{surfav:vel}~.
\end{eqnarray}
One important parameter characterizing the geometry in the transverse plane is the spatial eccentricity. During system expansion, anisotropic pressure gradients drive a stronger expansion along the shorter axis ($x$ axis), leading to a continuous reduction of the spatial eccentricity with time~\cite{Heinz:2004qz}. We define the time-dependent spatial eccentricity as~\cite{STAR:2014shf}
\begin{equation}
	\epsilon(t)=\frac{a^2(t)-b^2(t)}{a^2(t)+b^2(t)}.
\end{equation}
Fig.~\ref{fig:spacial eccentricity} shows the time evolution of the spatial eccentricity for different beam energies $\sqrt{s_{\rm NN}}$ using the parameters obtained from fitting the pion spectra. At early times, the system exhibits a sizable eccentricity reflecting the initial geometric deformation. As the system evolves, $\epsilon(t)$ decreases monotonically due to anisotropic collective expansion. The rate of this decrease is found to be stronger at higher collision energies, indicating more efficient pressure-driven expansion and a longer lifetime of the medium. Experimentally, the spatial eccentricity of the particle-emitting source at kinetic freeze-out can be extracted using azimuthally differential two-pion Hanbury Brown--Twiss (HBT) interferometry, through the oscillations of the HBT radii with respect to the reaction plane~\cite{Retiere:2003kf}. Using this method, the STAR Collaboration demonstrated that the freeze-out eccentricity remains finite but is significantly reduced compared to its initial value, and decreases systematically with increasing beam energy in Au+Au collisions~\cite{STAR:2014shf}.
\begin{table}[htbp]
	\centering
	\caption{Fitted model parameters at different collision energies.}
	\resizebox{\columnwidth}{!}{%
		\begin{tabular}{c r r r r r r r r}
			\hline\hline
			$\sqrt{s_{\rm NN}}$ & $T_{\rm kin}$ & $A$ & $B$ & $v_\infty$ & $\Delta v$ & $z_0$ & $v_0$ & $t_{f}$ \\
			(GeV) & (GeV) & & & & & & & (fm) \\
			\hline
			7.7  & 0.125 & 0.250 & 0.004 & 0.45 & 0.1 & 1.476 & 0.310 & 9.0  \\
			11.5 & 0.127 & 0.360 & 0.009 & 0.46 & 0.1 & 0.989 & 0.320 & 9.5  \\
			19.6 & 0.130 & 0.450 & 0.015 & 0.47 & 0.1 & 0.580 & 0.343 & 10.0 \\
			27   & 0.132 & 0.470 & 0.017 & 0.48 & 0.1 & 0.421 & 0.350 & 10.4 \\
			39   & 0.133 & 0.500 & 0.019 & 0.50 & 0.1 & 0.292 & 0.400 & 10.5 \\
			\hline\hline
		\end{tabular}%
	}
	\label{tab:model_params}
\end{table}

\begin{figure}[H]
	\centering
	\includegraphics[width=1\linewidth]{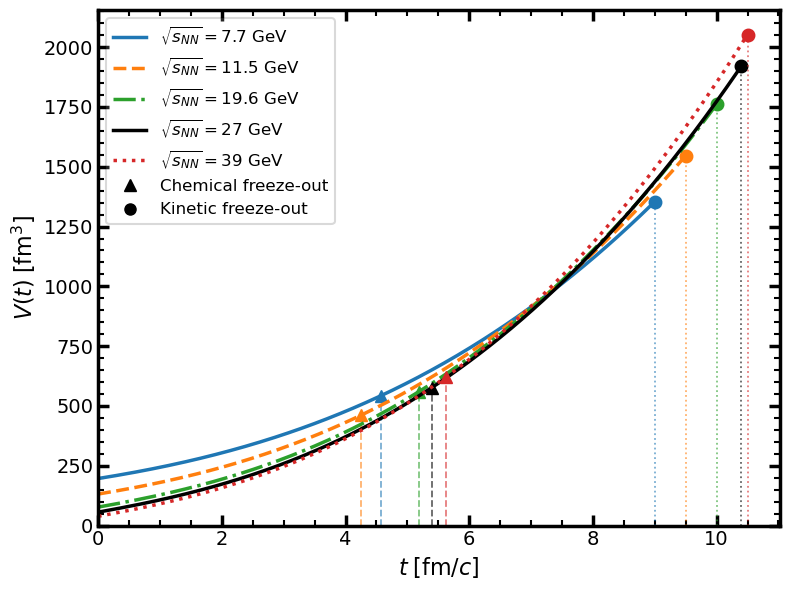}
	\caption{(Color online) Time evolution of the volume $V(t)$ of the expanding elliptic fire-cylinder for different collision energies $\sqrt{s_{\rm NN}}$.}
	\label{fig:volume evolution}
\end{figure}
\begin{table}[htbp]
	\centering
\caption{Surface velocity in the transverse plane compared with the STAR~\cite{STAR:2017sal} analysis for the centrality class $40$--$50\%$.}
	\begin{tabular}{c r r r r r}
		\hline\hline
		$\sqrt{s_{\rm NN}}$ (GeV) & 7.7 & $11.5$ & $19.6$ & $27$ & $39$  \\
		\hline
		STAR~\cite{STAR:2017sal}   & 0.507 & 0.582 & 0.614 & 0.648 & 0.690  \\
		This work $\langle v_{B}\rangle$  & 0.403 & 0.445 & 0.465 & 0.476 & 0.497  \\
		\hline\hline
	\end{tabular}
	\label{tab:vel}
\end{table}

\begin{figure}[H]
	\centering
	\begin{subfigure}{0.48\textwidth}
		\includegraphics[width=\linewidth]{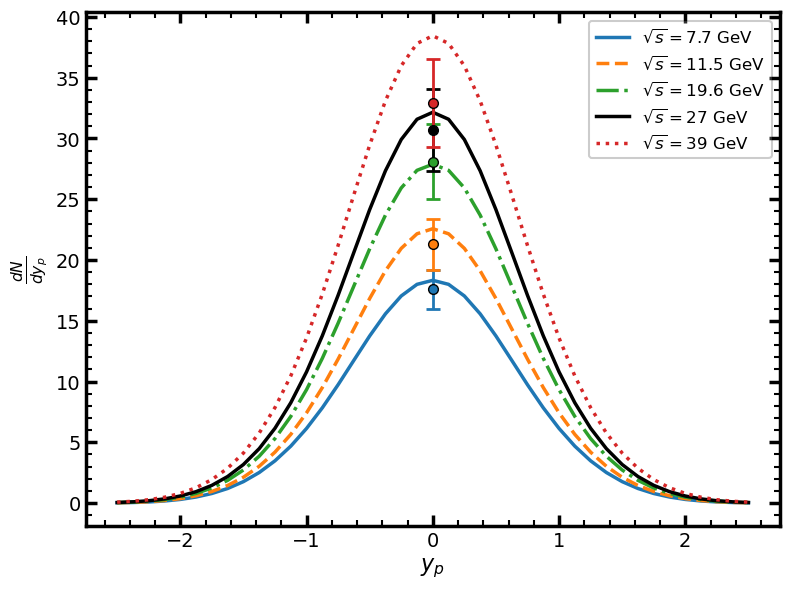}
	\end{subfigure}
	\hfill
	\begin{subfigure}{0.48\textwidth}
		\includegraphics[width=\linewidth]{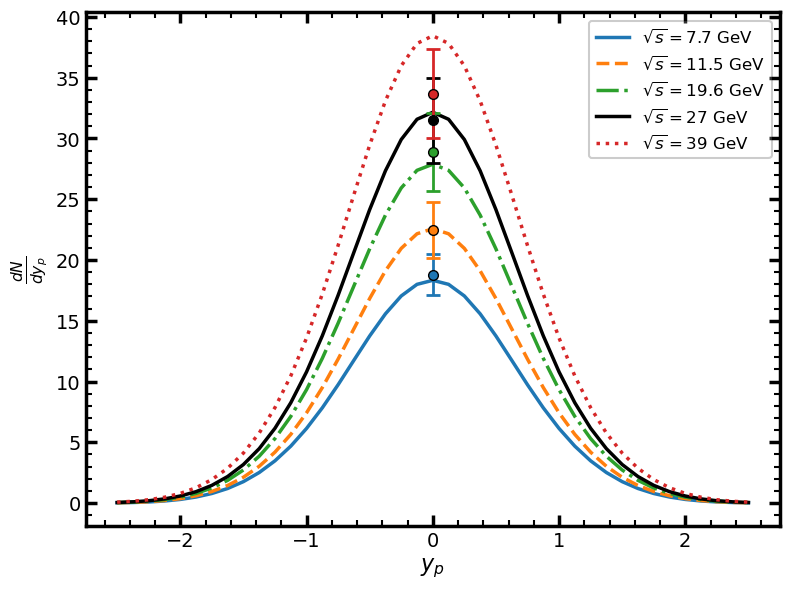}
	\end{subfigure}
	\caption{ (Color Online) Rapidity distribution of $\pi^{+}$ (upper panel) and $\pi^{-}$ (lower panel) for mid-central collisions at different beam energies $\sqrt{s_{\rm NN}}$ compared with the experimental data ~\cite{STAR:2017sal} for the centrality class ($40$--$50\%$) at mid-rapidity.}
	\label{fig:pion spetra rapidity dependance}
\end{figure}
\begin{figure*}[t]
	\centering
	\includegraphics[width=0.4\linewidth]{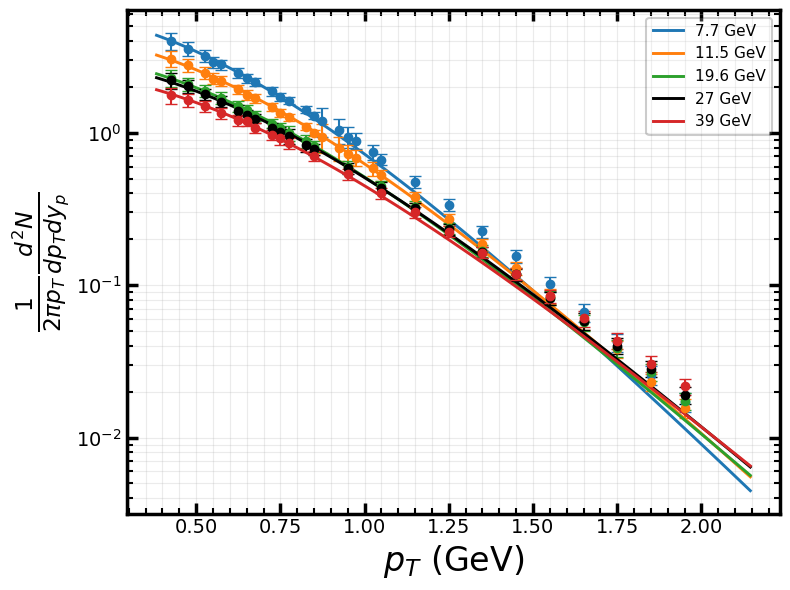}
	\includegraphics[width=0.4\linewidth]{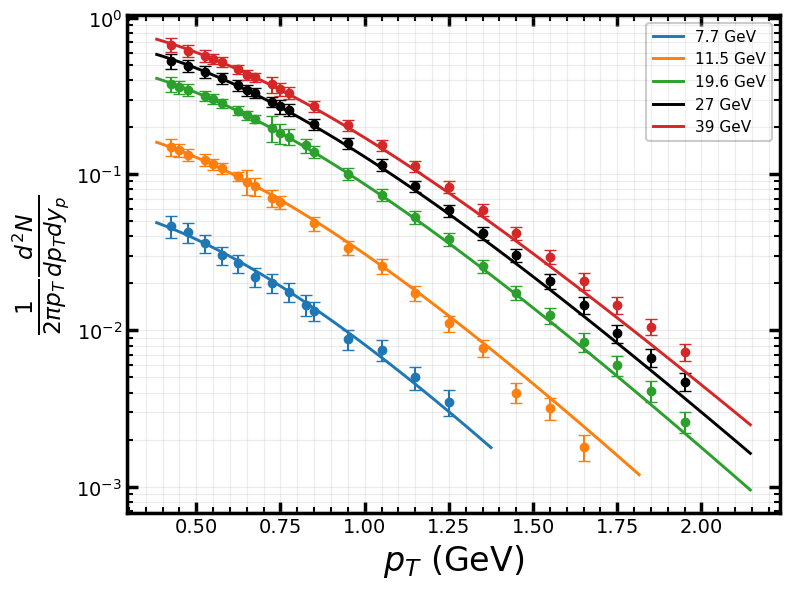}
	\includegraphics[width=0.4\linewidth]{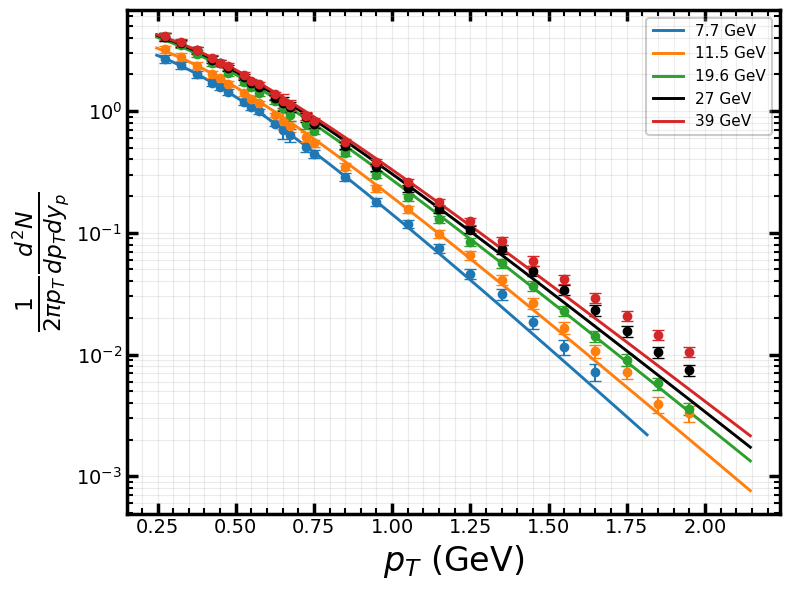}
	\includegraphics[width=0.4\linewidth]{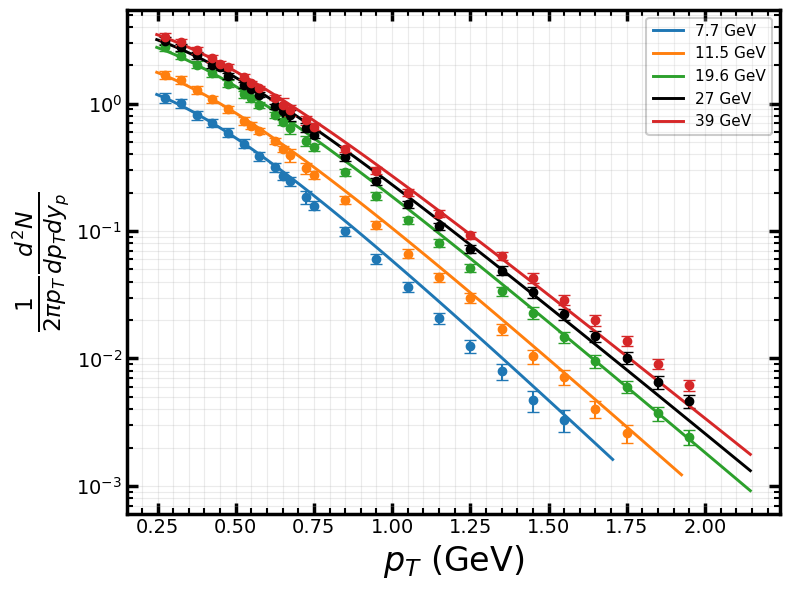}
	\caption{(Color online) Calculated $p_T$ spectra of protons ($p$) (top-left), anti-protons ($\bar{p}$) (top-right), positively charged kaons ($K^{+}$) (bottom-left), and negatively charged kaons ($K^{-}$) (bottom-right) at mid-rapidity compared with the data from~\cite{STAR:2017sal} in the centrality class ($40$--$50\%$).}
	\label{fig:proton-kaon-results}
\end{figure*}

 We note that the STAR Collaboration in Ref.~\cite{STAR:2014shf} reports the kinetic freeze-out eccentricity (see Fig.~28 of Ref.~\cite{STAR:2014shf}) for the centrality class $10$–$30$\%, whereas in the present work we analyze mid-central collisions. The magnitude reported by the STAR Collaboration lies in the range $\sim 0.16$–$0.10$ for $\sqrt{s_{\rm NN}} = 7.7$–$39$ GeV, while in our case it falls within $\sim 0.34$–$0.26$. For more central collisions (i.e., smaller centrality percentages), one generally expects a lower eccentricity due to the reduced initial spatial eccentricity. Nevertheless, interesting features emerge in the comparison performed in Ref.~\cite{STAR:2014shf} between the results of Ultrarelativistic Quantum Molecular Dynamics (UrQMD)~\cite{Bass:1998ca} and hydrodynamic simulations. While the predictions in both the approaches exhibit a decreasing trend of eccentricity with increasing beam energy for a given centrality class, the UrQMD results show better agreement with the experimental data than the $(2+1)$-dimensional hydrodynamic simulations. The hydrodynamic calculations tend to overestimate the eccentricity and are quite sensitive to the choice of equation of state employed in the simulations. In the present study, we restrict ourselves to a qualitative interpretation, which indicates a decreasing trend of freeze-out eccentricity with increasing beam energy. This conclusion should be viewed in light of the simplicity of our analysis and the difference in centrality selection compared to Ref.~\cite{STAR:2014shf}.  Analysis in the paper~\cite{STAR:2014shf} also shows an increasing medium lifetime with the beam energy. The qualitative agreement between the present results and experimental observations~\cite{STAR:2014shf} supports the consistency of our dynamical description of the fire-cylinder expansion.

 Similarly, using the chemical freeze-out analysis of the STAR collaboration, one can obtain the system volume at chemical freeze-out~\cite{STAR:2017sal} (see appendix~\ref{volcal}). These chemical freeze-out volumes, based on STAR collaboration data (chemical freeze-out potential $\mu_{\rm ch}$ and temperature $T_{\rm ch}$) and kinetic freeze-out volumes, based on this fire-cylinder model, are denoted by triangle and circle points, respectively, in Fig.~\ref{fig:volume evolution}. Along with the graphical representation, those chemical and kinetic freeze-out volumes and times are tabulated in Table~\ref{tab:volume data}.
\begin{table}[htbp]
	\centering
	\caption{Volume at kinetic and chemical freeze-out.}
	\label{tab:volume data}
	\begin{tabular}{c c  c c c}
		\hline\hline
		$\sqrt{s_{\rm NN}}$ & $t_{ch}$ & $V_{ch}$  & $t_{kin}$ & $V_{kin}$\\
		(GeV) & $fm$ & ${fm}^3$ & $fm$ & ${fm}^3$ \\
		\hline
		7.7 & 4.56 & 541.74  & 9.0 &  1354.32 \\
		11.5 & 4.25 & 461.69 & 9.5 &  1546.56  \\
		19.6 & 5.18 & 559.25 & 10.0 & 1761.15 \\
		27 & 5.40 & 578.76  & 10.4 & 1921.02 \\
		39  & 5.62 & 620.01 & 10.5 &  2052.67 \\
		\hline\hline
	\end{tabular}
\end{table}
Observation of $V_{\rm ch}(t_{\rm ch})<V_{\rm kin}(t_{f})$ and $t_{\rm ch}<t_{f}$ can be considered as a good consistency check of the hadronic phase evolution of the present model. Another consistency check for the model is provided by obtaining the rapidity spectrum and comparing it with the experimental mid-rapidity yield~\cite{STAR:2017sal} $\left(\frac{dN}{dy_{p}}\right)_{y_{p}=0}$. Rapidity distribution in low energy HICs remains Gaussian~\cite{Harabasz:2020sei}, we also observe a similar behavior in Fig.~\ref{fig:pion spetra rapidity dependance}.  

Once extracted as a result of fitting with the pion spectra, the resulting freeze-out parameter sets ($v_{\infty},\Delta v, v_{0}, A, B, t_{f}, T_{\rm kin}$) is applied consistently to the species $K^{\pm}$ and $p~(\bar{p})$ without further modification of the collective dynamics. This strategy ensures a unified description of soft hadron production, where differences between particle species arise solely from their masses and chemical potentials. In Fig.~\ref{fig:proton-kaon-results}, we show the $p_{T}$ spectra of protons and kaons at mid-rapidity. In this procedure, $\mu_p$ ($\mu_{\bar{p}}$) and $\mu_{K^{+}}$ ($\mu_{K^{-}}$) is the only adjustable parameters for particle (anti-particle). 
The resulting proton and anti-proton spectra are shown in the top-panel of Fig.~\ref{fig:proton-kaon-results}. The model provides a simultaneous description of the spectral shape, indicating that the inclusion of a finite chemical potential is essential to get consistent trends for proton (anti-proton) yields, particularly at lower beam energies where baryon stopping is significant~\cite{Cleymans:1998fq,Andronic:2005yp}.
The extracted values of $\mu_p$ (or $\mu_{\bar{p}}$) can be seen in Table~\ref{tab:mu_values}. 
The transverse-momentum spectra of $K^{+}$ and $K^{-}$ are shown in bottom panel of Fig.~\ref{fig:proton-kaon-results}. Our model also provides reasonable results for both the spectral shapes of $K^{+}$ and $K^{-}$ over the measured $p_T$ range with chemical potential value of the kaons given in Table~\ref{tab:mu_values}. The slight underestimation of the high $p_{T}$ region of the proton and kaon spectra may suggest that these particles freeze-out under conditions different from those of pions~\cite{Chatterjee:2013yga,Chatterjee:2014lfa}, i.e., the freeze-out parameter set  ($v_{\infty},\Delta v, v_{0}, A, B, t_{f}, T_{\rm kin}$) for them can be chosen differently to have a better match with the high $p_{T}$ data. However, this ambitious method is not pursued in the present work, and a common freeze-out condition is assumed for simplicity. At this juncture we wish to mention that the measured value of the chemical potentials for the protons and the kaons at kinetic freeze-out are directly sensitive to the baryon ($\mu^{B}_{\rm ch}$) and strange ($\mu^{S}_{\rm ch}$) chemical potentials at chemical freeze-out for the Hadron-resonance gas system~\cite{STAR:2017sal}. To see how the chemical potentials at the kinetic freeze-out and chemical freeze-out are related one see that the particle multiplicity (ignoring the hadron decay contributions) can be written as~\cite{Harabasz:2022rdt}
\begin{eqnarray}
N_{i}&=&\frac{g_{i}}{(2\pi)^3} \int_{\sigma_{\rm ch}} \int_{p} f_{i}(\mu^{i}_{\rm ch}, T_{\rm ch})~ p^\mu \frac{d^3\vec{p}}{E}  d\sigma_\mu \nonumber\\
&=& \frac{g_{i}}{(2\pi)^3} \int_{\sigma_{\rm kin}} \int_{p} f_{i}(\mu^{i}_{\rm kin}, T_{\rm kin})~ p^\mu \frac{d^3\vec{p}}{E}  d\sigma_\mu	\label{ch-kin}
\end{eqnarray}
where $i$ denotes the particle species. $\sigma_{\rm ch}$ and $\sigma_{\rm kin}=\sigma$ are respectively the chemical and kinetic freeze-out hyper surfaces. At chemical freeze-out the system is specified by three independent chemical potentials corresponding to three conserved quantum numbers  (baryon $B_{i}$, charge $Q_{i}$ and strangeness $S_{i}$) with chemical potential of species $i$ given by $\mu^{i}_{\rm ch}=\mu^{B}_{\rm ch}B_{i}+\mu^{Q}_{\rm ch}Q_{i}+\mu^{S}_{\rm ch}S_{i}$. From Eq.~\eqref{ch-kin} one obtains the following relation for local number densities, 
\begin{eqnarray}
\frac{n^{i}_{\rm ch}}{n^{j}_{\rm ch}}=\frac{n^{i}_{\rm kin}}{n^{j}_{\rm kin}}, \implies \mu^{i}_{\rm kin}-\mu^{j}_{\rm kin}\simeq\frac{T_{\rm kin}}{T_{\rm ch}} (\mu^{i}_{\rm ch}-\mu^{j}_{\rm ch}),	\label{dens}
\end{eqnarray}
where to get the last equality we assume Boltzmann distribution. Using Eq.~\eqref{dens} for protons and kaons we obtain $\mu_{p,K^{+}}-\mu_{\bar{p},K^{-}}\simeq2\frac{T_{\rm kin}}{T_{\rm ch}} (\mu^{B,S}_{\rm ch})$. From this relation one can check that the difference between chemical potentials between particle-antiparticle pair is lesser at kinetic freeze-out than at chemical freeze-out (see Table~\eqref{tab:mu_values}).

\begin{figure*}[htbp]
	\centering
	\includegraphics[width=0.4\linewidth]{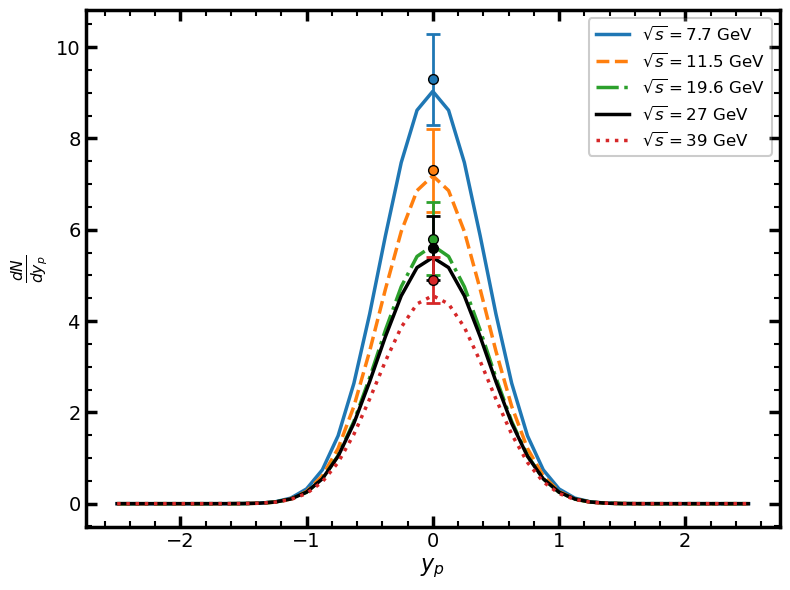}
	\includegraphics[width=0.4\linewidth]{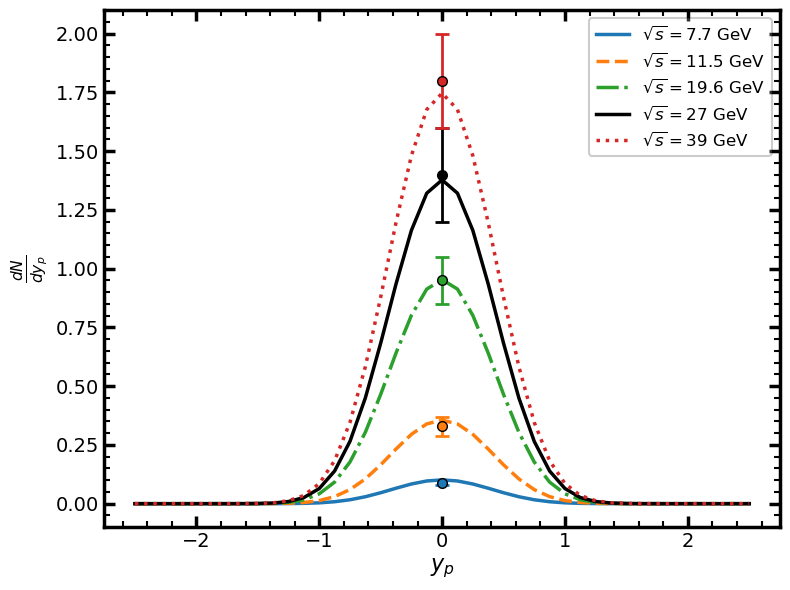}
	\includegraphics[width=0.4\linewidth]{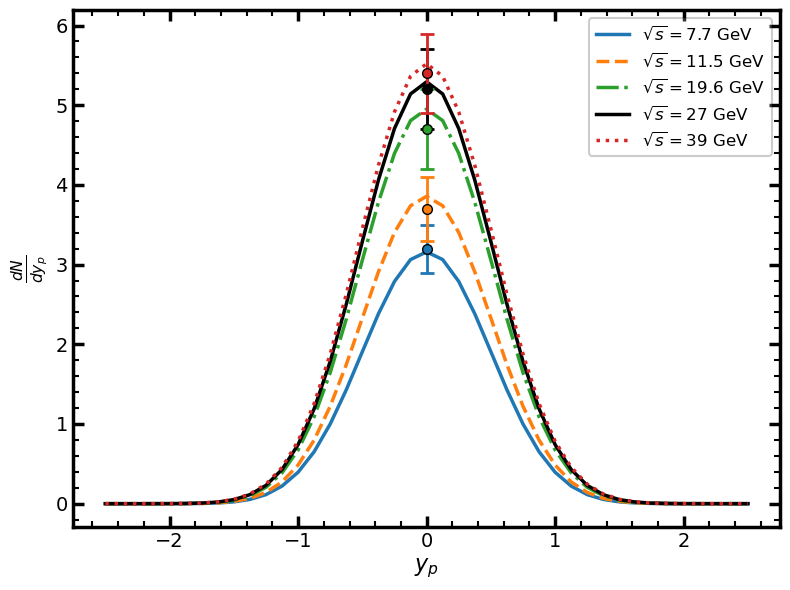}
	\includegraphics[width=0.4\linewidth]{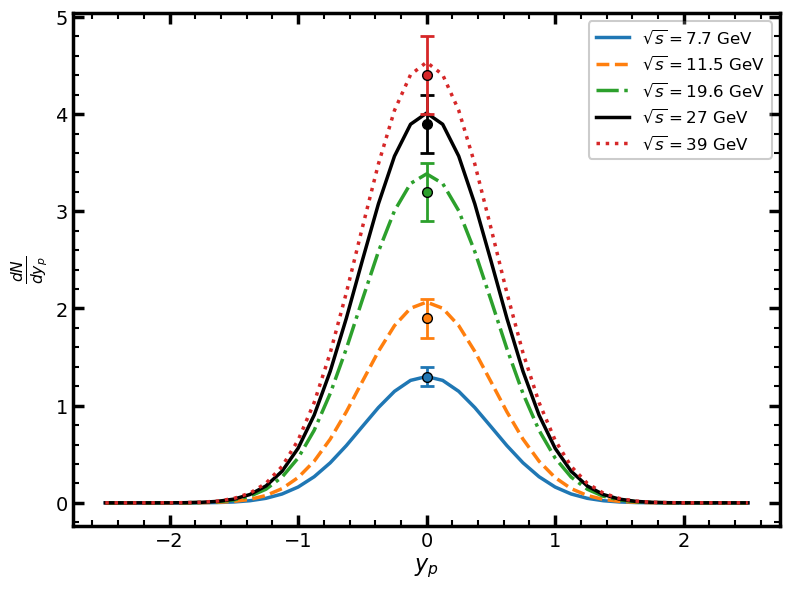}
	\caption{Rapidity distribution of $p$ (top-left), $\bar{p}$ (top-right), $K^{+}$ (bottom-left) and $K^{-}$ (bottom-right) for mid-central collisions at different beam energies $\sqrt{s_{\rm NN}}$ compared with the data from~\cite{STAR:2017sal} in the centrality class ($40$--$50\%$).}
	\label{fig:rapidity dependance}
\end{figure*}

The beam-energy dependence of the rapidity distributions shows that the overall particle yield increases with increasing $\sqrt{s_{\rm NN}}$, as expected from the larger initial entropy production at higher collision energies~\cite{Cleymans:1999st,Andronic:2017pug}. We present the rapidity distribution of the $p$ ($\bar{p}$) and $K^{+}$ ($K^{-}$) in Fig.~\ref{fig:rapidity dependance}. The distribution appears Gaussian-like, similar to the pion rapidity distribution presented previously. In contrast to the longitudinal boost invariance scenario where rapidity distributions have a plateau near mid-rapidity~\cite{Bjorken:1982qr,Heinz:2004qz}, low energy HICs show a Gaussian distribution~\cite{NA49:2002pzu,HADES:2020ver}. The Gaussian rapidity distributions have been obtained theoretically from non-boost invariant blast wave models in Refs.~\cite{Rode:2018hlj,Rode:2020vhu}.
	These Gaussian profiles have been observed in low energy HICs at CERN and GSI~\cite{NA49:2002pzu,HADES:2020ver} which even exists till $\sqrt{s_{\rm NN}}=200$ GeV~\cite{Ouerdane:2002gm,PhysRevLett.93.102301,PhysRevLett.94.162301}. We do not attempt a direct quantitative comparison with those results due to differences in the colliding system, collision energy, and centrality selection. However, we have shown the data of STAR collaboration~\cite{STAR:2017sal} measured at mid-rapidity in Fig.~\ref{fig:rapidity dependance}, which are well fitted by the peak value of a Gaussian-like rapidity distribution, obtained from the present model. 
	
\begin{figure}[H]
	\centering
	\includegraphics[width=1\linewidth]{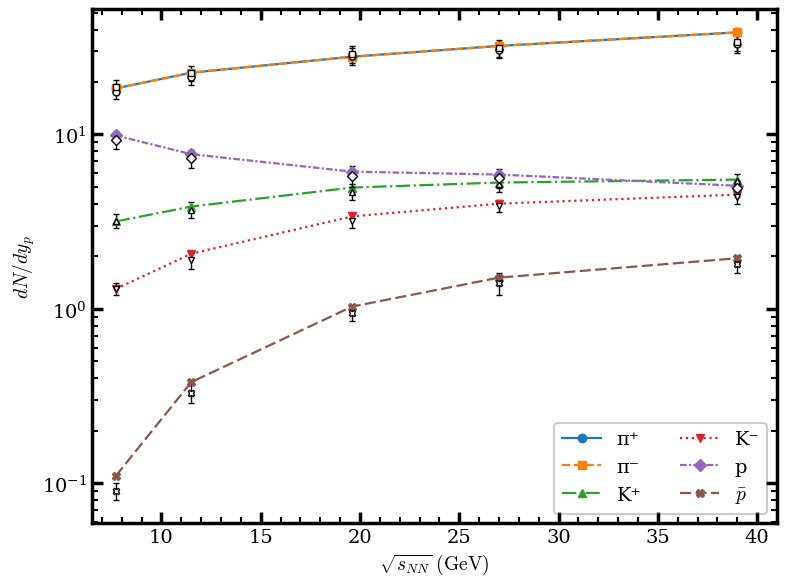}
	\caption{Mid-rapidity particle yields $dN/dy_{p}$ as a function of $\sqrt{s_{\rm NN}}$ for $\pi^{\pm}$, $K^{\pm}$, $p$, and $\bar{p}$. Symbols with error bars represent experimental data~\cite{STAR:2017sal} in the centrality class ($40$--$50\%$), while lines denote model calculations.}
	\label{fig:yield}
\end{figure}

	By compiling Fig.~\ref{fig:pion spetra rapidity dependance} and \ref{fig:rapidity dependance}, we can get the beam energy dependence of mid-rapidity particle yield $dN/dy_p$ as a function of $\sqrt{s_{\rm NN}}$, for $\pi^{+}$, $\pi^{-}$, $K^{+}$, $K^{-}$, $p$, and $\bar{p}$. The results are shown in Fig.~\ref{fig:yield} and Table~\ref{tab:yield_comparison}, where experimental measurements are compared with the corresponding model calculations. The model calculations reproduce the overall energy dependence of the measured yields for all the particle species considered. In particular, the approximate equality of $\pi^{+}$ and $\pi^{-}$ yields indicates a negligible or minor charge chemical potential for the system. 
	Quantitative differences between the model and experimental data remain within the experimental uncertainties, indicating that the essential features of particle production across the explored energy range are reasonably well described.
\begin{table}[htbp]
\centering
		\caption{Chemical potentials $\mu$ (GeV) for proton ($p$), kaon$^{+}$ ($K^{+}$), anti-proton ($\bar{p}$), and kaon$^{-}$ ($K^{-}$) at kinetic freeze-out compared with the STAR collaboration~\cite{STAR:2017sal} chemical freeze-out parameters ($T_{\rm ch}$, $\mu_{Q}=0$, $\mu^B_{\rm ch}$, and $\mu^S_{\rm ch}$) for the  Grand Canonical Ensemble yield (GCEY) fits in the centrality class ($40$--$60\%$). Errors in
			parenthesis are systematic uncertainties. All values except $\sqrt{s_{\rm NN}}$ are in MeV.}
	 \label{tab:mu_values}
	\begin{tabular}{c c c c c c c c}
			\hline\hline
		$\sqrt{s_{\rm NN}}$ & $T_{\rm ch}$ & $\mu^B_{\rm ch}$ & $\mu^S_{\rm ch}$ & $\mu_{p}$ & $\mu_{\bar{p}}$ & $\mu_{K^{+}}$& $\mu_{K^{-}}$ \\
		\hline
		7.7  & 145.5 (2.7)  & 357.8 (12.0) & 82.2 (7.0) & 323 & -238 & -40 & -151 \\
		11.5 & 157.9 (3.7)  &259.2 (12.6) & 62.5 (7.6)  & 263 & -119 & -45 & -124 \\
		19.6 & 162.2 (3.5)  &159.4 (9.8) & 40.1 (6.3)  & 197 & -35 & -46 & -96 \\
		27   & 165.5 (3.5)  &127.5 (8.9) & 34.9 (5.7)  & 167 & -14 & -61 & -97 \\
		39   & 163.5 (3.5)  &86.8 (9.9)  & 23.2 (6.7)  & 124 & -4 & -80 & -106 \\
		\hline\hline
	\end{tabular}
\end{table}

\begin{table*}[htbp]
	\centering
	\caption{Comparison of experimental in the centrality class ($40$--$50\%$)~\cite{STAR:2017sal} and model-calculated mid-rapidity particle yields $dN/dy_{p}$
		for Au+Au collisions at different center-of-mass energies $\sqrt{s_{\rm NN}}$.
		Experimental uncertainties are statistical only.}
	\label{tab:yield_comparison}
	\begin{tabular}{c c c c c c c}
	\hline\hline
		$\sqrt{s_{NN}}$ (GeV)
		& Particle
		& Exp.
		& Model
		& Particle
		& Exp. & Model \\
		\hline
		
		\multirow{6}{*}{7.7}
		& $\pi^{+}$ & $17.6 \pm 1.6$ & 18.37 & $\pi^{-}$ & $18.8 \pm 1.7$ & 18.37 \\
		& $K^{+}$   & $3.2 \pm 0.3$  & 3.17  & $K^{-}$   & $1.3 \pm 0.1$  & 1.30  \\
		& $p$       & $9.3 \pm 1.0$  & 9.03  & $\bar{p}$ & $0.09 \pm 0.01$ & 0.10 \\
		\hline
		
		\multirow{6}{*}{11.5}
		& $\pi^{+}$ & $21.3 \pm 2.1$ & 22.62 & $\pi^{-}$ & $22.5 \pm 2.3$ & 22.62 \\
		& $K^{+}$   & $3.7 \pm 0.4$  & 3.86  & $K^{-}$   & $1.9 \pm 0.2$  & 2.07  \\
		& $p$       & $7.3 \pm 0.9$  & 7.18  & $\bar{p}$ & $0.33 \pm 0.04$ & 0.36 \\
		\hline
		
		\multirow{6}{*}{19.6}
		& $\pi^{+}$ & $28.1 \pm 3.1$ & 27.93 & $\pi^{-}$ & $28.9 \pm 3.2$ & 27.93 \\
		& $K^{+}$   & $4.7 \pm 0.5$  & 4.96  & $K^{-}$   & $3.2 \pm 0.3$  & 3.39  \\
		& $p$       & $5.8 \pm 0.8$  & 5.66  & $\bar{p}$ & $0.95 \pm 0.10$ & 0.95 \\
		\hline
		
		\multirow{6}{*}{27}
		& $\pi^{+}$ & $30.7 \pm 3.4$ & 32.21 & $\pi^{-}$ & $31.5 \pm 3.5$ & 32.21 \\
		& $K^{+}$   & $5.2 \pm 0.5$  & 5.30  & $K^{-}$   & $3.9 \pm 0.3$  & 4.01  \\
		& $p$       & $5.6 \pm 0.7$  & 5.39  & $\bar{p}$ & $1.4 \pm 0.2$  & 1.39 \\
		\hline
		
		\multirow{6}{*}{39}
		& $\pi^{+}$ & $32.9 \pm 3.6$ & 38.51 & $\pi^{-}$ & $33.7 \pm 3.7$ & 38.51 \\
		& $K^{+}$   & $5.4 \pm 0.5$  & 5.51  & $K^{-}$   & $4.4 \pm 0.4$  & 4.52  \\
		& $p$       & $4.9 \pm 0.5$  & 4.56  & $\bar{p}$ & $1.8 \pm 0.2$  & 1.81 \\
		
		\hline\hline
	\end{tabular}
\end{table*}

\subsection{Elliptic flow results}
\label{subsec:v2_results}
\begin{figure*}[htbp]
	\centering
	\includegraphics[width=0.4\linewidth]{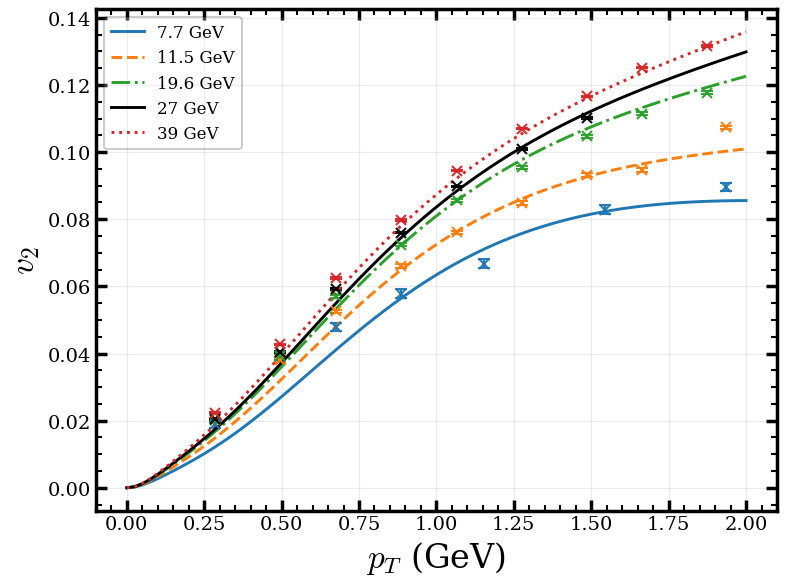}
	\includegraphics[width=0.4\linewidth]{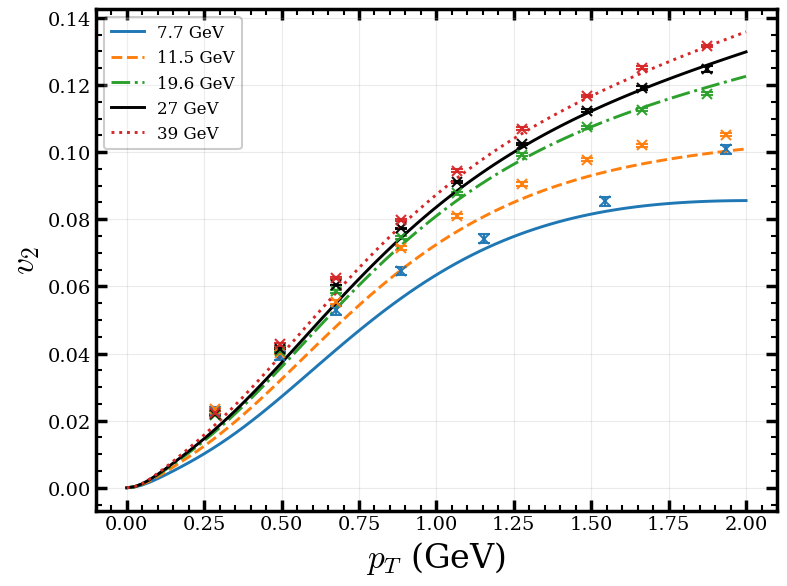}
	\includegraphics[width=0.4\linewidth]{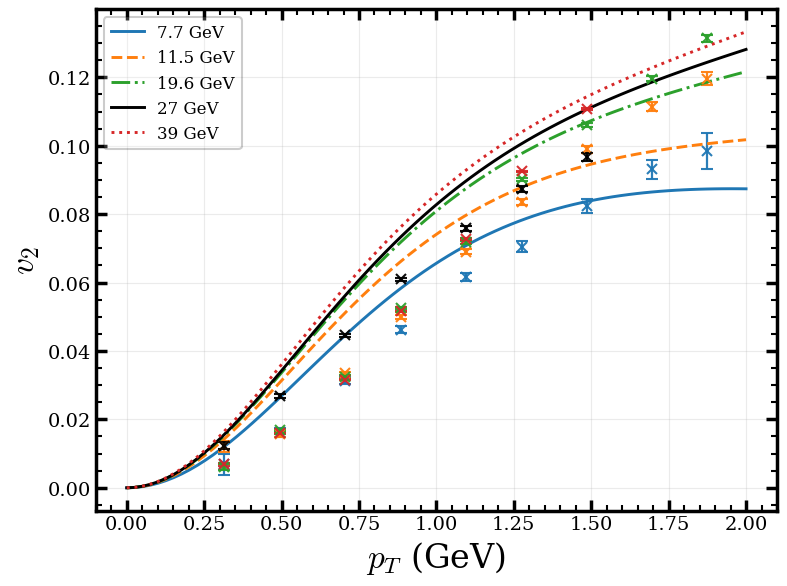}
	\includegraphics[width=0.4\linewidth]{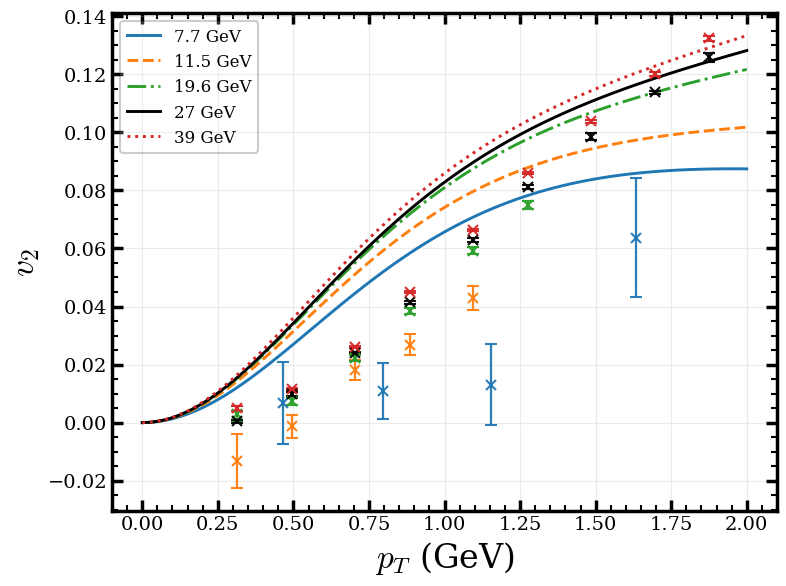}
	\includegraphics[width=0.4\linewidth]{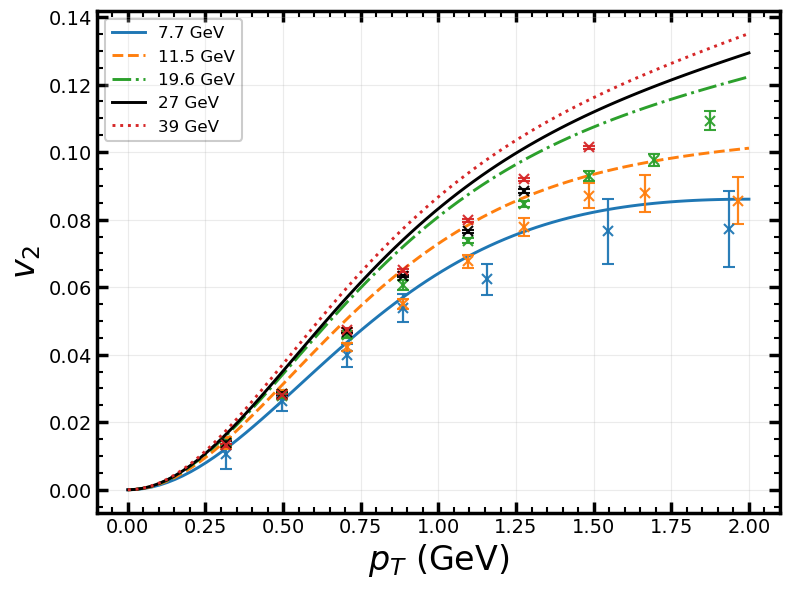}
	\includegraphics[width=0.4\linewidth]{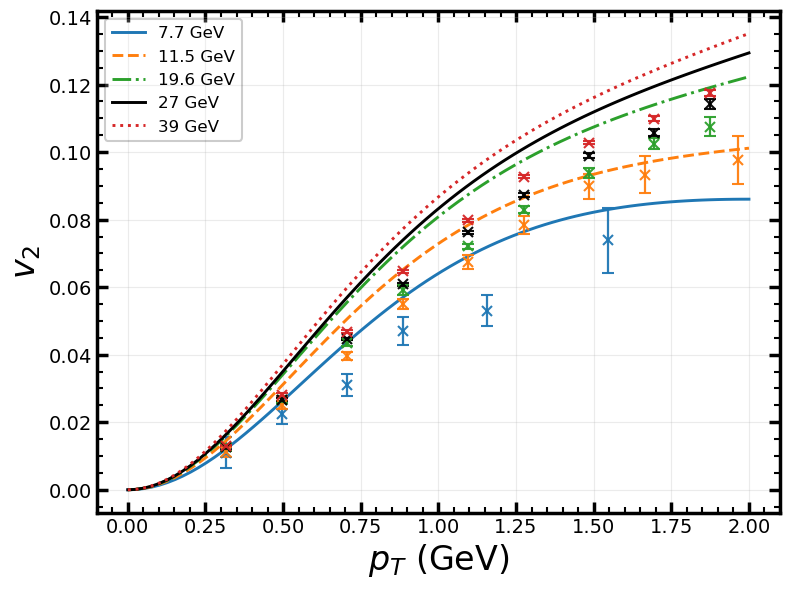}
	\caption{$p_{T}$ dependence of the elliptic flow $v_2(p_T)$ for $\pi^{+}$ (top-left), $\pi^{-}$ (top-right), $p$ (middle-left), $\bar{p}$ (middle-right), $K^{+}$ (bottom-left), and $K^{-}$ (bottom-right) obtained using the same expansion parameters as in the $\pi$ sector and chemical potentials from Table~\ref{tab:mu_values} is compared with the STAR data~\cite{STAR:2013ayu} of $0-80\%$ centrality.}
	\label{fig:v2_all}
\end{figure*}

In addition to the transverse-momentum spectra, the present framework provides a simultaneous description of the elliptic flow coefficient $v_2$. Using the same fire-cylinder geometry and hydrodynamic expansion parameters that were fixed from the fits to the identified-hadron $p_T$ spectra, the calculated $v_2(p_T)$ exhibits a qualitative agreement with the measured trends for pions, kaons, protons, and their corresponding anti-particles.
In Fig.~\ref{fig:v2_all},
we present the results of $p_{T}$-dependent elliptic flow at mid-rapidity along with the corresponding experimental data from STAR collaboration~\cite{STAR:2013ayu}. The predicted elliptic flow shows the characteristic rise with transverse momentum at low $p_T$, followed by a gradual saturation toward higher momenta. This behavior reflects the buildup of collective anisotropic flow driven by the initial spatial eccentricity and the subsequent pressure gradients, as originally proposed in Ref.~\cite{Ollitrault:1992bk} and extensively discussed in later theoretical and experimental studies~\cite{STAR:2013ayu,Voloshin:2008dg,Heinz:2013th}. Within the present model, the magnitude and shape of $v_2(p_T)$ are governed by the transverse expansion parameter $\Delta v$ and $B$, which control the strength and time development of the collective flow. In our calibration we fix $\Delta v$ but increase $B$ with increasing $\sqrt{s_{\rm NN}}$ (see Table~\ref{tab:model_params}) which through Eqs.~\ref{eq:adot} and \ref{eq:bdot} imply a rapid development of anisotropic velocity part $\Delta v$ in time (see Fig.~\ref{fig:velocity_profile}). The increment in $B$ makes the elliptic flow higher at higher $\sqrt{s_{\rm NN}}$, which is reflected for all the considered particles in Fig.~\ref{fig:v2_all}. We observe excellent quantitative agreement between our results and the experimental data for pions. For pions and kaons, the qualitative trend and ranking for different $\sqrt{s}$ are not very different in our model calculation and experiments. However, for protons and antiprotons, an appreciable difference is observed. Since the number ($\frac{dN}{dy_{p}}|_{y_{p}=0}$) of anti-protons is quite lower than that of protons and other mesons, its results may need a statistical fluctuation physics (event by event dynamics) consideration along with the thermalized statistical physics. The negative value of the elliptic flow for anti-protons obtained in the experiments for $\sqrt{s_{\rm NN}}$ remains elusive. As discussed in Ref.~\cite{Retiere:2003kf}, higher radial flow (corresponding to $v_{\infty}$ in the present study) can make the $v_{2}(p_{T})$ for heavier particles negative at low $p_{T}$. Therefore, one may need to choose a larger radial boost for the anti-protons, as done in Ref.~\cite{Sun:2014rda}, to better align the results with the experimental data. However, we do not pursue it here in view of the simplicity of the model.

\section{SUMMARY}\label{sec:sum}
In summary, we study the transverse momentum spectra and elliptic flow of light hadrons produced in the mid-central Au+Au collisions at the Relativistic Heavy Ion Collider for relatively low beam energies. The system formed in such collisions is modeled as an expanding cylinder with an elliptic cross section to account for the geometry of off-central collisions. The longitudinal and transverse expansion rates are parameterized using five unknown constants. The observables are calculated using the Cooper–Frye freeze-out prescription at a constant laboratory time, assuming a single-particle thermal equilibrium distribution function.

The mid-rapidity pion spectra are used to constrain the five expansion parameters, along with the kinetic freeze-out temperature and freeze-out time. The resulting parameters characterizing the transverse eccentricity and freeze-out time are found to be qualitatively consistent with results from identical-particle interferometry measurements, which probe the system at kinetic freeze-out. Assuming a single freeze-out hypersurface and using the parameters fixed from the pion spectra, we subsequently predict the spectra and elliptic flow of $K^{\pm}$, $p$, and $\bar{p}$. The predicted results are compared directly with experimental data to assess the quantitative agreement of the model. By introducing particle-specific chemical potentials for $K^{\pm}$, $p$, and $\bar{p}$ as additional fitting parameters, we are able to get consistent trends of their transverse momentum spectra. The elliptic flow of $\pi^{\pm}$, $K^{\pm}$, $p$, and $\bar{p}$ obtained with the same set of model parameters shows overall qualitative agreement with experimental observations.

This work offers an alternative blast-wave–like parametrization for describing heavy-ion collisions at lower beam energies. For simplicity, several effects such as resonance decays and baryon stopping, which may influence the observed spectra and flow, are not included in the present analysis and will be addressed in future studies. We plan to extend the model to additional beam energies and centrality classes, and to investigate directed flow, higher-order flow observables such as triangular flow, as well as the effects of electromagnetic fields. Furthermore, once the soft-sector observables are well constrained, the model can be employed as a background framework to study penetrating probes such as heavy-quark diffusion and electromagnetic radiation.

\section{ACKNOWLEDGMENT}
This work was supported in part by the Board of Research in Nuclear Sciences (BRNS) and the Department of Atomic Energy (DAE), Government of India, with Grant Nos. 57/14/01/2024-BRNS/313 (A.R. and S.G.) and the Ministry of Education, Government of India (A.D.). The authors thank Sathwik HS, Ankit Kumar Panda, and Bhagyarathi Sahoo for fruitful discussions.


\onecolumngrid
\section{Appendix}
\label{appe1}
\subsection{Shape and expansion of the transverse ellipse}
\label{subsec:shape}

In this appendix, we see the direction of the transverse velocity in relation to the periphery of the ellipse. Before that, it is worth mentioning that to obey the causality constraint we should have,
\begin{eqnarray}
	&&v_{x}^{2}+ v_{y}^{2} \leq 1	\nonumber\\
	\implies && \dot{b}^{2} \left(\frac{r \cos\theta}{r_{B}}\right)^{2} + \dot{a}^{2} \left(\frac{r \sin\theta}{r_{B}}\right)^{2}\leq 1~.
\end{eqnarray}
From which the sufficient condition becomes that the expansion velocity on the perimeter of the ellipse at the freeze-out time should be less than one, i.e., $\dot{b}^{2}(t_{f})+ \dot{a}^{2}(t_{f})\leq 1$.
\begin{figure}[H]
	\centering
	\begin{subfigure}{0.3\textwidth}
		\includegraphics[width=\linewidth]{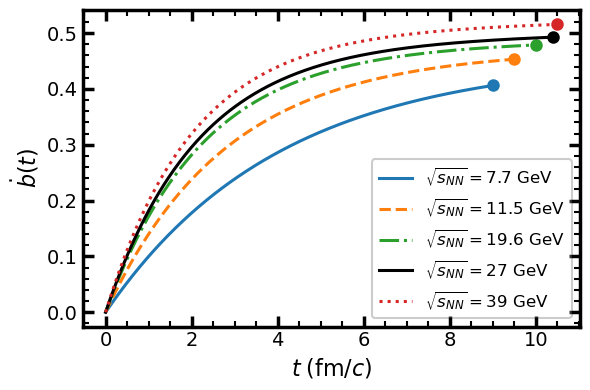}
	\end{subfigure}
	\hfill
	\begin{subfigure}{0.3\textwidth}
		\includegraphics[width=\linewidth]{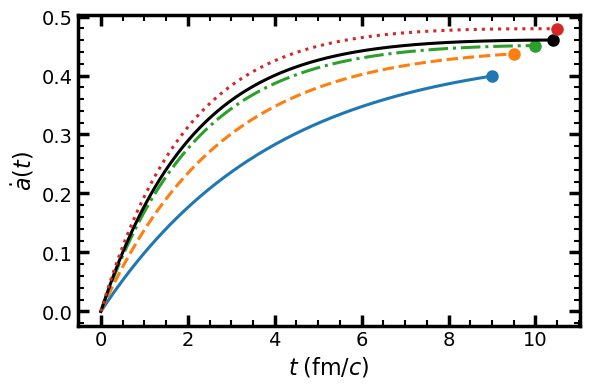}
	\end{subfigure}
	\hfill
	\begin{subfigure}{0.3\textwidth}
		\includegraphics[width=\linewidth]{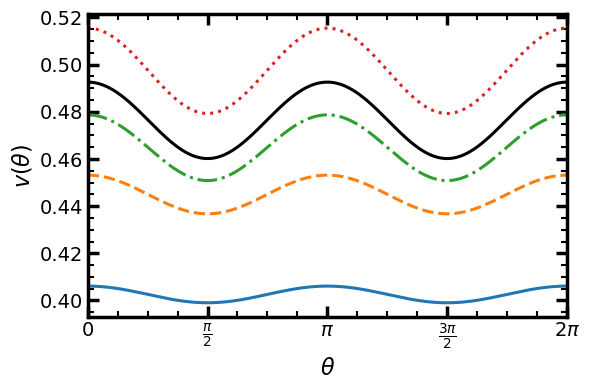}
	\end{subfigure}
	\caption{Time evolution and angular dependence of the transverse expansion velocity of the fire-cylinder for different collision energies $\sqrt{s_{NN}}$.}
	\label{fig:velocity_profile}
\end{figure}

\begin{figure}[H]
	\centering
	\begin{subfigure}{0.19\textwidth}
		\includegraphics[width=\linewidth]{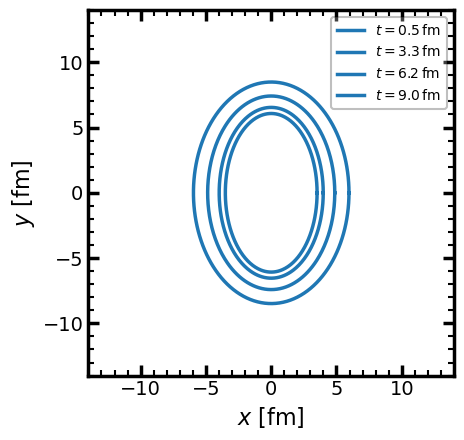}
		\caption{$7.7$ GeV}
	\end{subfigure}\hfill
	\begin{subfigure}{0.19\textwidth}
		\includegraphics[width=\linewidth]{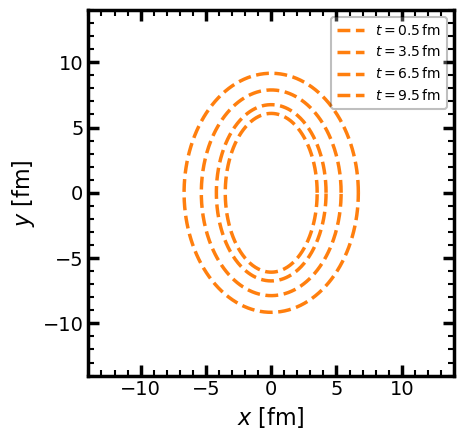}
		\caption{$11.5$ GeV}
	\end{subfigure}\hfill
	\begin{subfigure}{0.19\textwidth}
		\includegraphics[width=\linewidth]{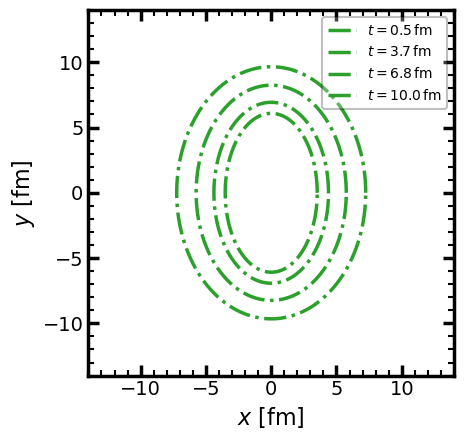}
		\caption{$19.6$ GeV}
	\end{subfigure}\hfill
	\begin{subfigure}{0.19\textwidth}
		\includegraphics[width=\linewidth]{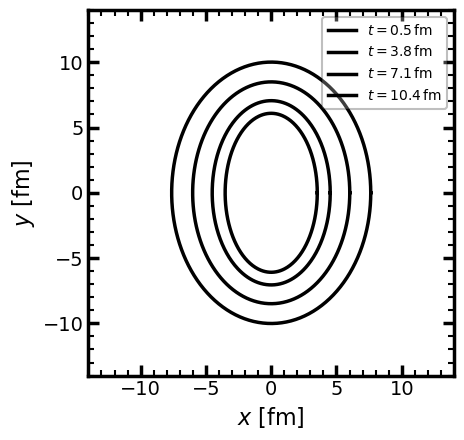}
		\caption{$27$ GeV}
	\end{subfigure}\hfill
	\begin{subfigure}{0.19\textwidth}
		\includegraphics[width=\linewidth]{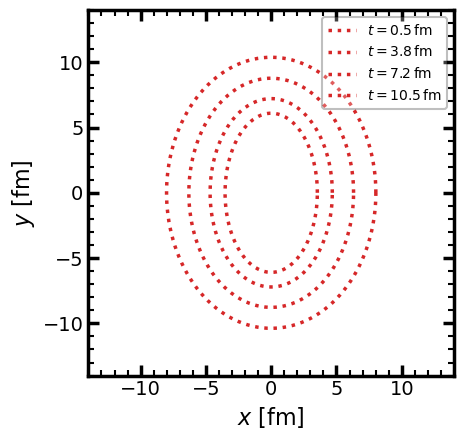}
		\caption{$39$ GeV}
	\end{subfigure}
	\caption{Snapshots of the elliptical evolution of the expanding fire-cylinder at four different times.}
	\label{fig:ellipse evolution}
\end{figure}
To check the direction of the transverse velocity, we write down the tangent vector to the ellipse as,
\begin{eqnarray}
	\vec{n}_{t}(\xi, \theta)= \frac{\partial \vec{r}_{T}}{\partial \theta}&=&\frac{\partial}{\partial \theta}~ (\lambda(t) \sinh\xi \cos\theta~ \hat{i} + \lambda(t) \cosh\xi \sin\theta ~\hat{j})\nonumber\\
	&=&-	\lambda(t) \sinh\xi \sin\theta~ \hat{i} + \lambda(t) \cosh\xi \cos\theta ~\hat{j}~.
\end{eqnarray}
Taking the dot product of the transverse velocity $\vec{v}_{T}$ with $\vec{n}_{t}$ we have,
\begin{eqnarray}
	\vec{n}_{t}(\xi, \theta)\cdot \vec{v}_{T}&=&(-\lambda(t) \sinh\xi \sin\theta~ \hat{i} + \lambda(t) \cosh\xi \cos\theta ~\hat{j}) \cdot \left(\frac{r}{r_{B}}\dot{b}\cos\theta~ \hat{i} + \frac{r}{r_{B}}\dot{a}\sin\theta~ \hat{j}\right)\nonumber\\
	&=& -\lambda(t) \frac{r}{r_{B}}\sin\theta\cos\theta ~(\dot{b}\sinh\xi-\dot{a}\cosh\xi).\label{dirvel}
\end{eqnarray}
From Eq.~\eqref{dirvel} we realize that for the velocity to be perpendicular to the elliptical boundary of the system we need $\dot{a}(t)/\dot{b}(t)=\tanh \xi_{\rm max}(t)$ which in turn implies $\lambda(t)=$ constant.

In Fig.~\ref{fig:velocity_profile}, we display the expansion of the minor (along $x$), major axis (along $y$), and azimuthal variation of the velocity in the left, middle, and right panels, respectively. The expansion of the ellipse's axes is shown up to $t=t_{f} (\sqrt{s_{\rm NN}})$ for different beam energies. To display the azimuthal variation of the velocity, only the boundary ellipse at the freeze-out time $t=t_{f}(\sqrt{s_{\rm NN}})$ is considered. We observe that the rate of expansion increases with larger beam energies, as expected, because of the larger pressure gradients as the energy of the collisions increases. Similarly, one can observe a higher rate of expansion along the $x$ axis than along the $y$ axis, i.e., $\dot{b}>\dot{a}$. The right panel of Fig.~\ref{fig:velocity_profile} essentially clarifies the beam energy dependence of $v_{2}$ behavior observed in the Sec.~\ref{sec:results}. A lesser azimuthal variation of the expansion velocity with decreasing $\sqrt{s_{\rm NN}}$ makes $v_{2}$ smaller at lower $\sqrt{s_{\rm NN}}$. Fig.~\ref{fig:ellipse evolution} presents the snapshots of the medium in the transverse plane at different times. We can observe that the ellipses' eccentricities gradually decrease over time.

\begin{figure}[H]
	\centering
	\includegraphics[width=0.4\linewidth]{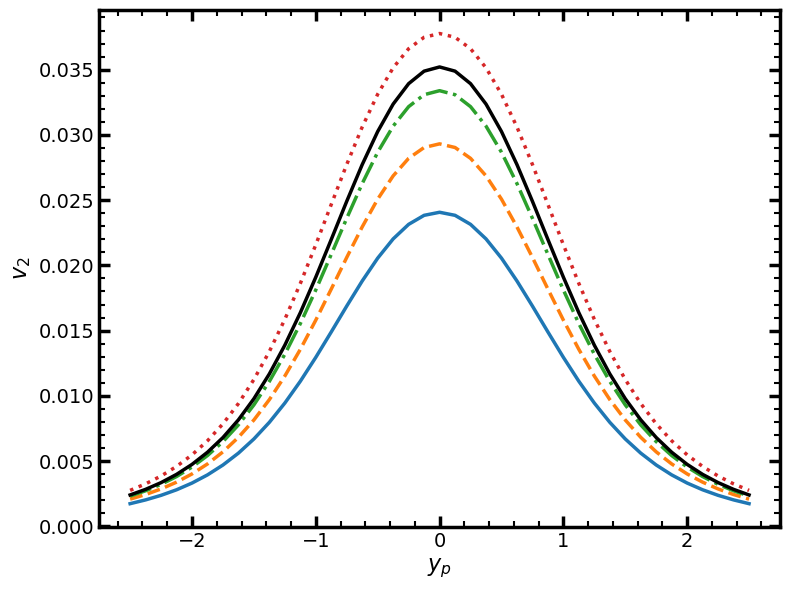}
	\includegraphics[width=0.4\linewidth]{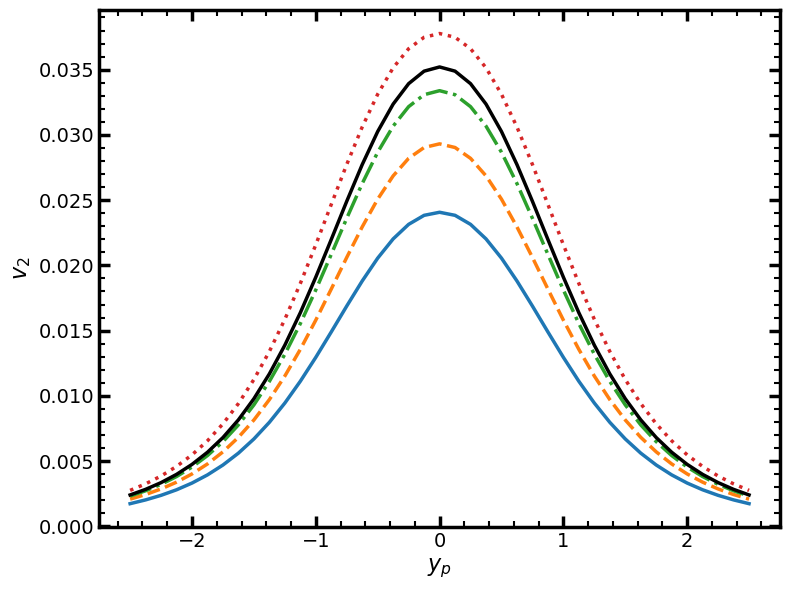}
	\includegraphics[width=0.4\linewidth]{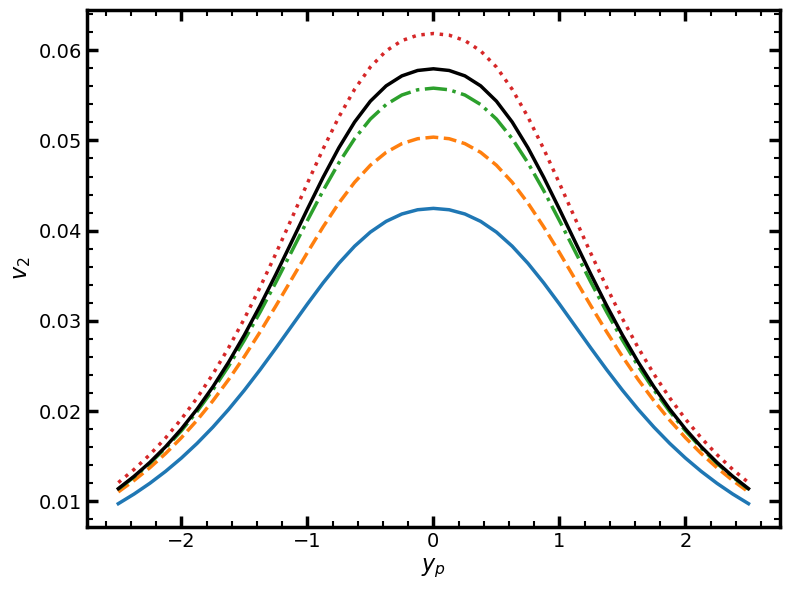}
	\includegraphics[width=0.4\linewidth]{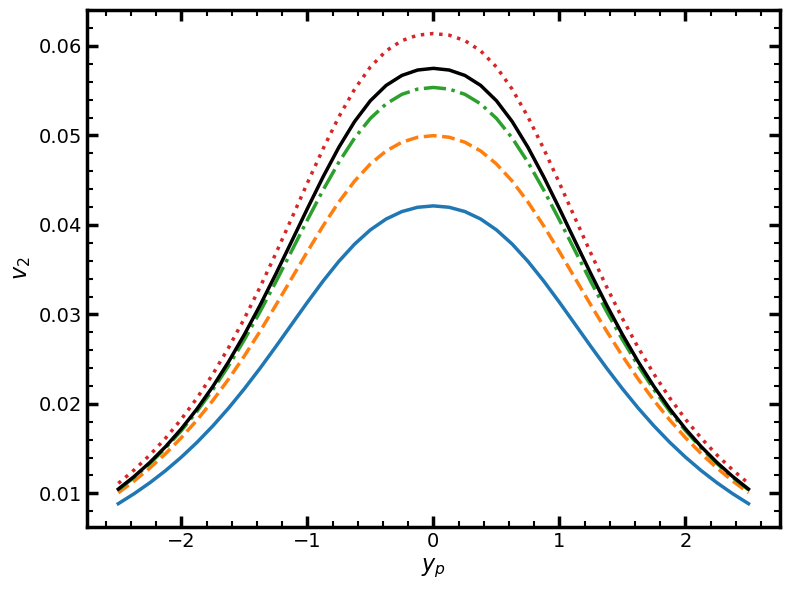}
	\includegraphics[width=0.4\linewidth]{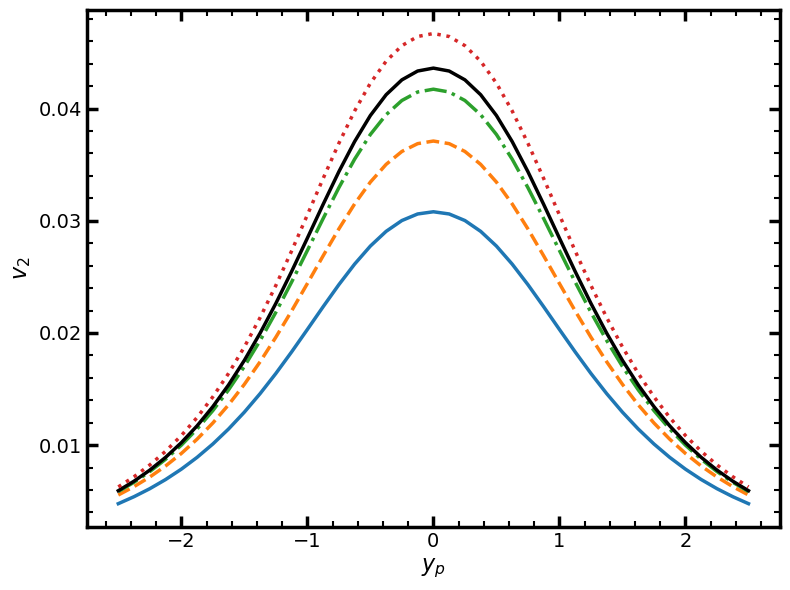}
	\includegraphics[width=0.4\linewidth]{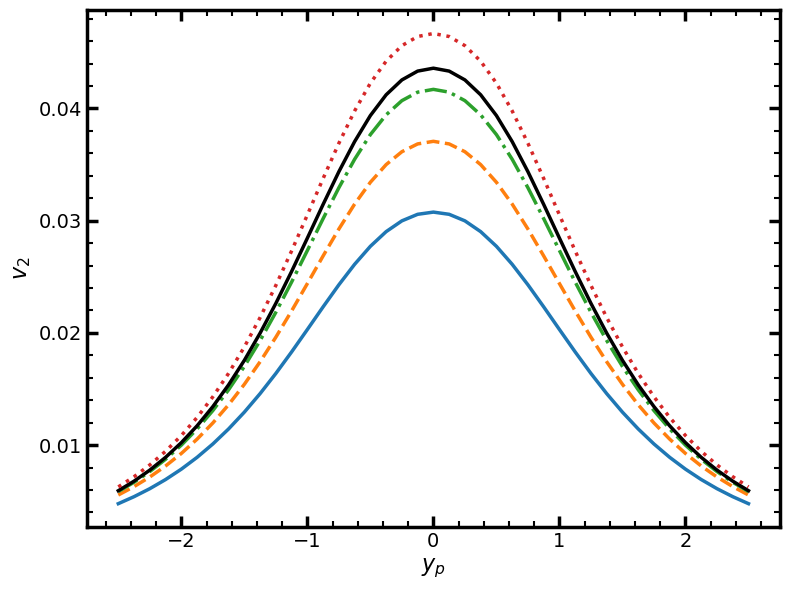}
	\caption{Rapidity dependence ($p_T$ integrated) of elliptic flow $v_2$ for mid-central collisions at different beam energies $\sqrt{s_{\rm NN}}$ is shown for $\pi^{+}$ (top-left), $\pi^{-}$ (top-right),  $p$ (middle-left), $\bar{p}$ (middle-right), $K^{+}$ (bottom-left) and $K^{-}$ (bottom-right).}
	\label{fig: rapidity dependancev2}
\end{figure}
\subsection{Rapidity dependence of elliptic flow}
The dependence of the collective flow on rapidity provides important insights into the longitudinal dynamics of the expanding medium produced in relativistic HICs. In Fig.~\ref{fig: rapidity dependancev2}, we present the momentum-rapidity dependence of the elliptic flow coefficient $v_2$ for mid-central collisions at different beam energies $\sqrt{s_{\rm NN}}$. As shown in Fig.~\ref{fig: rapidity dependancev2}, the elliptic flow exhibits a pronounced maximum around mid-rapidity ($y_{p} \approx 0$) and gradually decreases toward larger rapidities.
This behavior reflects the reduced azimuthal anisotropy in the particle production as one moves away from the region of mid-rapidity.

\subsection{Statistical Analysis}
\begin{figure}[H]
	\centering
	\includegraphics[width=0.4\linewidth]{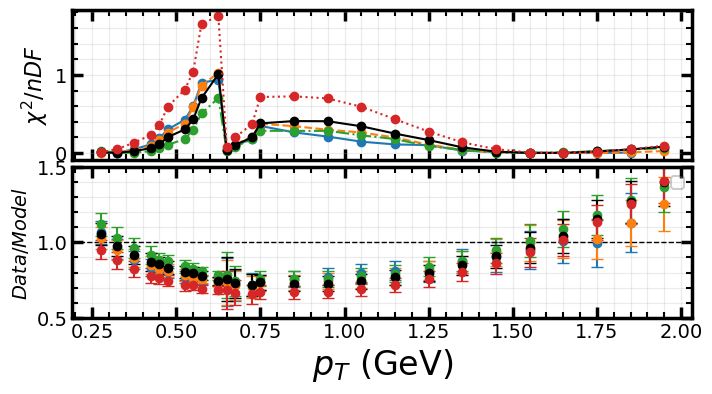}
	\includegraphics[width=0.4\linewidth]{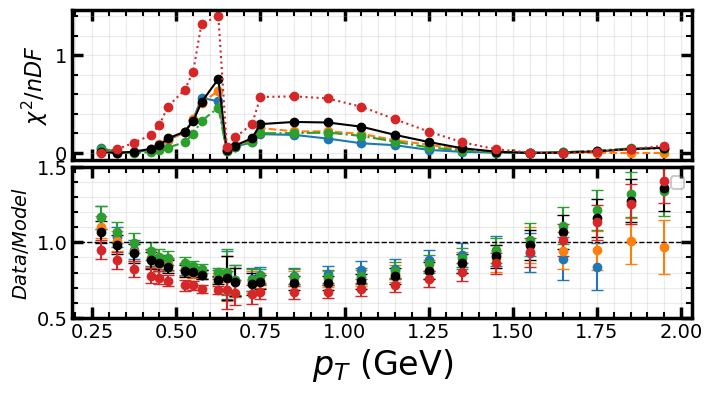}
	\includegraphics[width=0.4\linewidth]{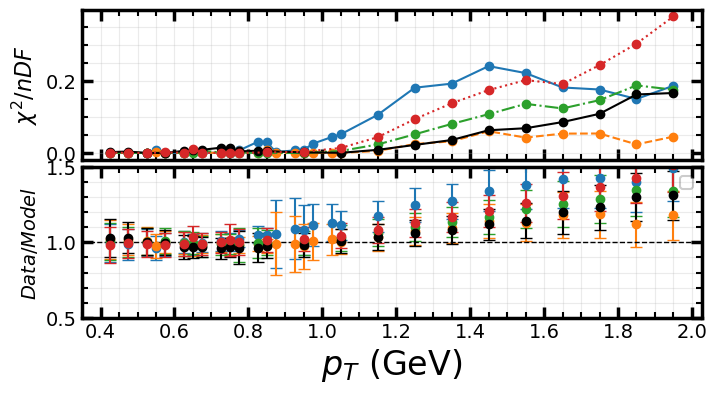}
	\includegraphics[width=0.4\linewidth]{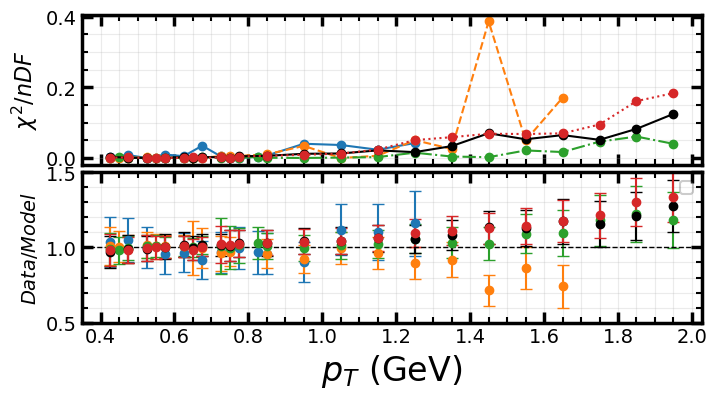}
	\includegraphics[width=0.4\linewidth]{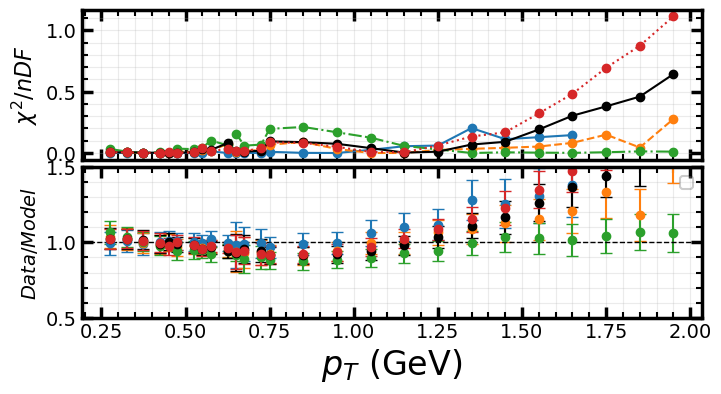}
	\includegraphics[width=0.4\linewidth]{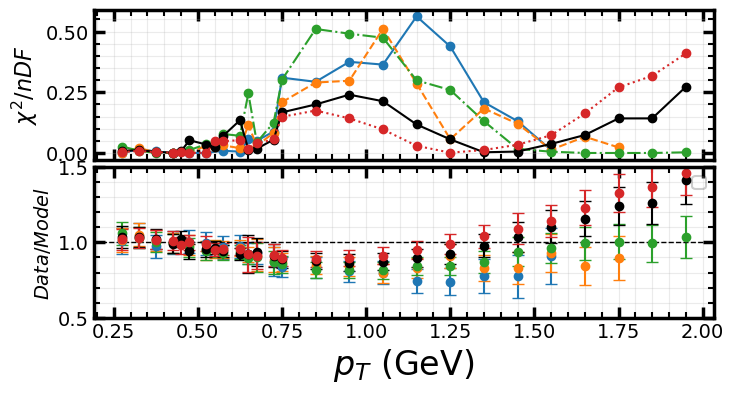}
	\caption{(Color online) Deviations between the measured and fitted transverse momentum ($p_T$) spectra for $\pi^{+}$ (top left), $\pi^{-}$ (top right), $p$ (middle left), $\bar{p}$ (middle right), $K^{+}$ (bottom left), and $K^{-}$ (bottom right).}
	\label{fig:proton-kaon-results1}
\end{figure}

The quality of the model description of the data was quantified using a $\chi^2$ minimization procedure. The $\chi^2$ function is defined as
\begin{equation}
	\chi^2 = \sum_{i=1}^{N}
	\frac{\left(D_i - M_i\right)^2}{\sigma_i^2},
\end{equation}
where $D_i$ denotes the experimental measurement, $M_i$ the corresponding model prediction, and $\sigma_i$ the total uncertainty of the data point. The larger uncertainty between the statistical ($\sigma_{i,\mathrm{stat}}$) and systematic ($\sigma_{i,\mathrm{syst}}$) uncertainties are used as $\sigma_{i}$. The number of degrees of freedom is defined as $\mathrm{nDF} = N - N_{\mathrm{par}}$,
where $N_{\mathrm{par}}$ is the number of fit parameters. For the pion spectra we take $N_{\rm par}=6$ ($A,B,v_{\infty}, v_{0}, t_{f}, T$) and for proton and kaons we take $N_{\rm par}=1$ (only chemical potential). The quantity $\chi^{2}/\rm{nDF}$ is displayed in the respective upper panels of the Fig.\ref{fig:proton-kaon-results1}. The quantity $R_i = \frac{D_i}{M_i}$ (denoted as Data/Model) is also shown in the lower panels with uncertainties propagated from the experimental errors as
$\Delta R_i = \frac{\sigma_i}{M_i}$.


\subsection{Volume calculation from STAR Data~\cite{STAR:2017sal}}\label{volcal}
The effective medium volume at chemical freeze-out is estimated from the measured particle multiplicity assuming thermal equilibrium and a Bose--Einstein (or Fermi-Dirac) distribution. To get the the particle yield we use
\begin{align}
	\Delta N&= \frac{dN}{dy_p}\Delta y_p ,\nonumber\\
	&=n(\mu_{ch},T_{ch}) \frac{dV}{dy_p}\Delta y_p ,\nonumber\\
	&=n(\mu_{ch},T_{ch}) V_{ch}\Delta y_p~,
\end{align}
where $V_{ch}\equiv \frac{dV}{dy_p}$ is the effective volume per rapidity and $n(\mu_{ch},T_{ch})$ is the thermal number density. For a particle with mass $m$, the number density is given by
\begin{equation}
	n(\mu_{\rm ch},T_{\rm ch}) = \frac{1}{2\pi^{2}}
	\int_{0}^{\infty}
	\frac{p^{2}\, dp}
	{\exp\left((E-\mu_{\rm ch})/T_{\rm ch}\right) \pm 1},
\end{equation}
where the single-particle energy is $E = \sqrt{\vec{p}^{2} + m^{2}}$ and the chemical potential $\mu_{\rm ch}=\mu^{B}_{\rm ch}B_{i}+\mu^{Q}_{\rm ch}Q_{i}+\mu^{S}_{\rm ch}S_{i}$. To estimate $V_{\rm ch}$, we take the chemical freeze-out value of baryon, strange, and charge chemical potentials and temperature provided in the analysis of~\cite{STAR:2017sal} (also listed in Table~\eqref{tab:mu_values}). This results in different $V_{\rm ch}$ for different particles.
The average volume $V_{\rm ch}$ of pion, proton, and kaon is shown in Fig.~\ref{fig:volume evolution}.

\twocolumngrid


\bibliographystyle{unsrturl}
\bibliography{1ref}
\end{document}